\documentclass[sigconf,screen,authorversion]{acmart}

\def\BibTeX{{\rm B\kern-.05em{\sc i\kern-.025em b}\kern-.08emT\kern-.1667em\lower.7ex\hbox{E}\kern-.125emX}}

\setcopyright{none} %
\renewcommand\footnotetextcopyrightpermission[1]{} % removes footnote with conf
\usepackage{multirow}

\begin{document}

%
% The "title" command has an optional parameter, allowing the author to define a "short title" to be used in page headers.
\title[Venue Analytics]{Venue Analytics: A Simple Alternative to Citation-Based Metrics}

%
% The "author" command and its associated commands are used to define the authors and their affiliations.
% Of note is the shared affiliation of the first two authors, and the "authornote" and "authornotemark" commands
% used to denote shared contribution to the research.

\author{Leonid Keselman}

\affiliation{%
  \institution{Carnegie Mellon University}
  \streetaddress{5000 Forbes Ave}
  \city{Pittsburgh}
  \state{PA}
  \postcode{15213}
}
\email{leonidk@cmu.edu}

%\author{Martial Hebert}
%\affiliation{%
%  \institution{Carnegie Mellon University}
%  \streetaddress{5000 Forbes Ave}
%  \city{Pittsburgh}
%  \state{PA}
%  \postcode{15213}
%}
%\email{hebert@ri.cmu.edu}
%
% By default, the full list of authors will be used in the page headers. Often, this list is too long, and will overlap
% other information printed in the page headers. This command allows the author to define a more concise list
% of authors' names for this purpose.
%\renewcommand{\shortauthors}{Keselman}

%
% The abstract is a short summary of the work to be presented in the article.
\begin{abstract}
We present a method for automatically organizing and evaluating the quality of different publishing venues in Computer Science. Since this method only requires paper publication data as its input, we can demonstrate our method on a large portion of the DBLP dataset, spanning 50 years, with millions of authors and thousands of publishing venues. By formulating venue authorship as a regression problem and targeting metrics of interest, we obtain venue scores for every conference and journal in our dataset. The obtained scores can also provide a per-year model of conference quality, showing how fields develop and change over time. Additionally, these venue scores can be used to evaluate individual academic authors and academic institutions. We show that using venue scores to evaluate both authors and institutions produces quantitative measures that are comparable to approaches using citations or peer assessment. In contrast to many other existing evaluation metrics, our use of large-scale, openly available data enables this approach to be repeatable and transparent. 

To help others build upon this work, all of our code and data is available at \url{https://github.com/leonidk/venue_scores}.
\end{abstract}

%
% The code below is generated by the tool at http://dl.acm.org/ccs.cfm.
% Please copy and paste the code instead of the example below.
%

%
% Keywords. The author(s) should pick words that accurately describe the work being
% presented. Separate the keywords with commas.
%\keywords{ranking,venues,conferences,nsf,dblp,regression,citation,metrics}

%
% A "teaser" image appears between the author and affiliation information and the body 
% of the document, and typically spans the page. 
\begin{teaserfigure}
    \includegraphics[width=0.195\textwidth]{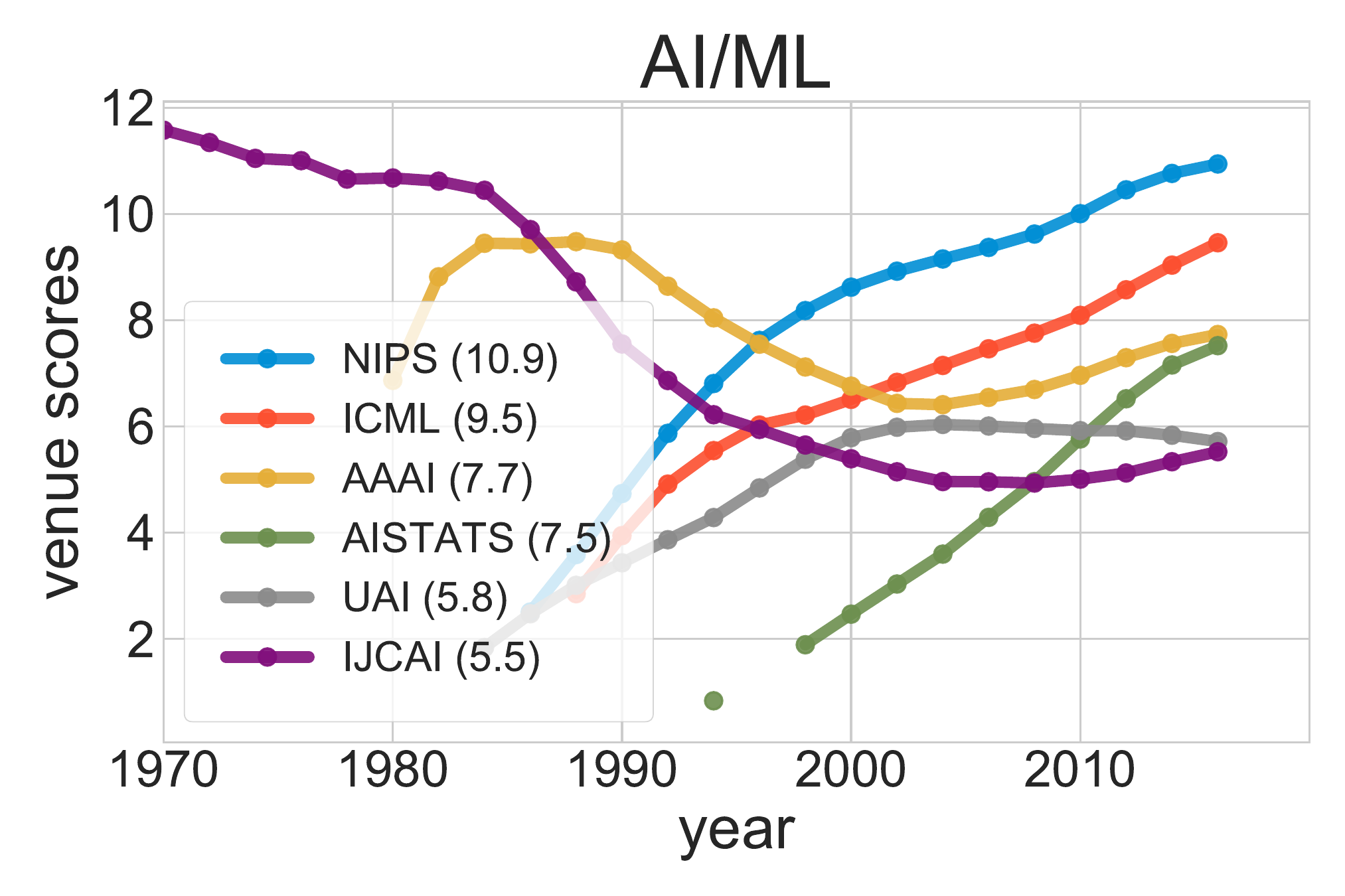}
    \includegraphics[width=0.195\textwidth]{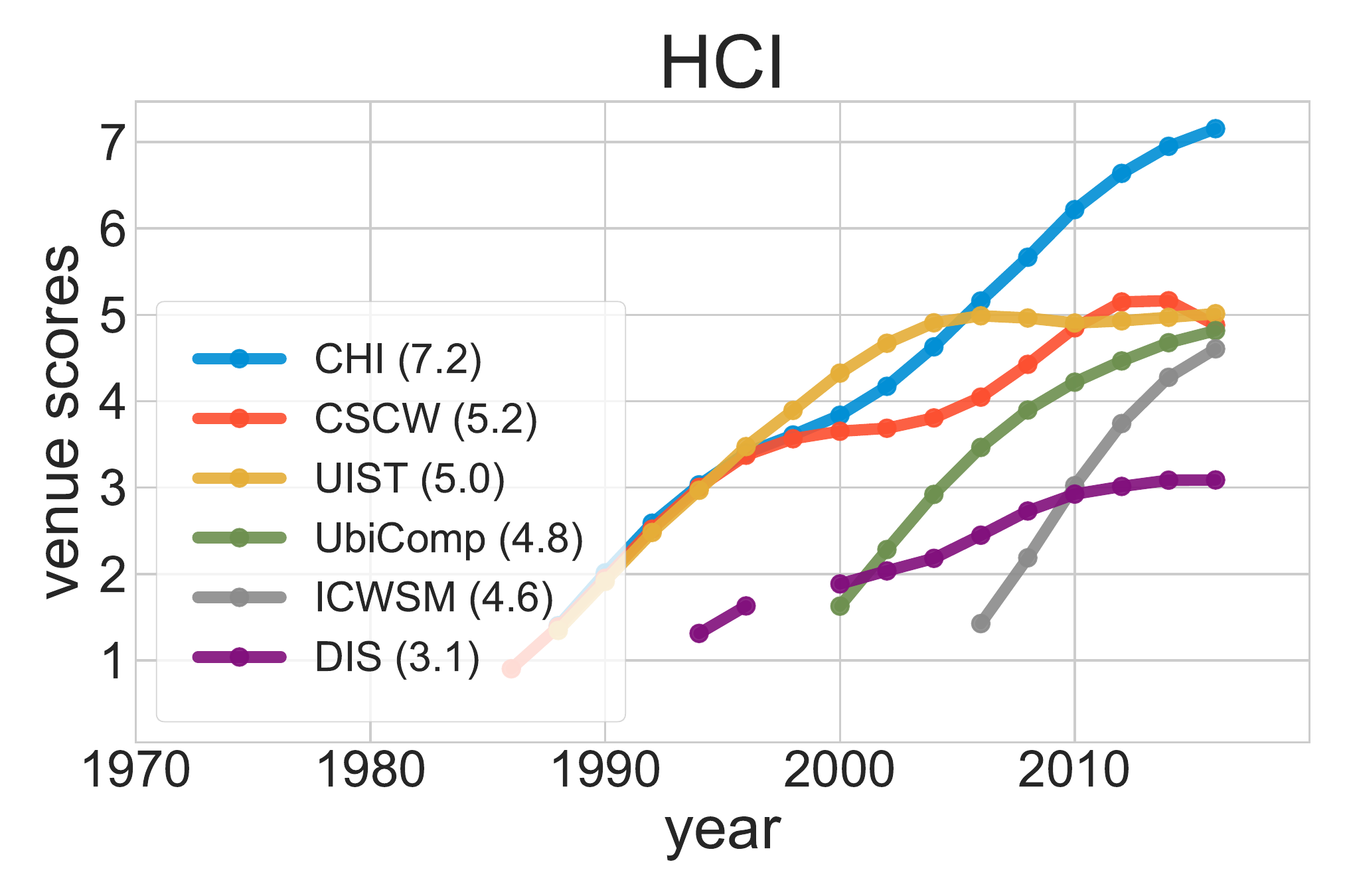}
    \includegraphics[width=0.195\textwidth]{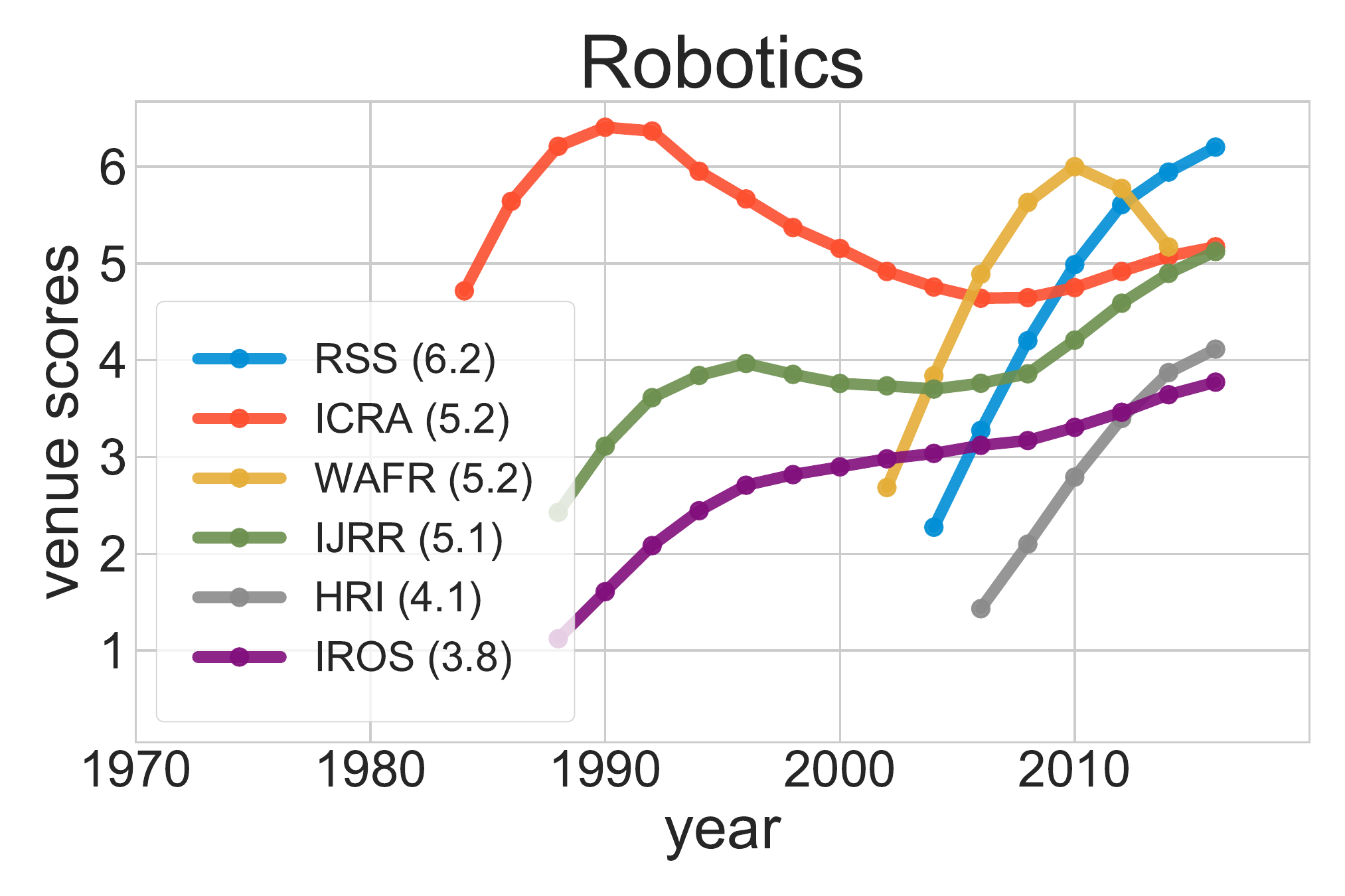}
    \includegraphics[width=0.195\textwidth]{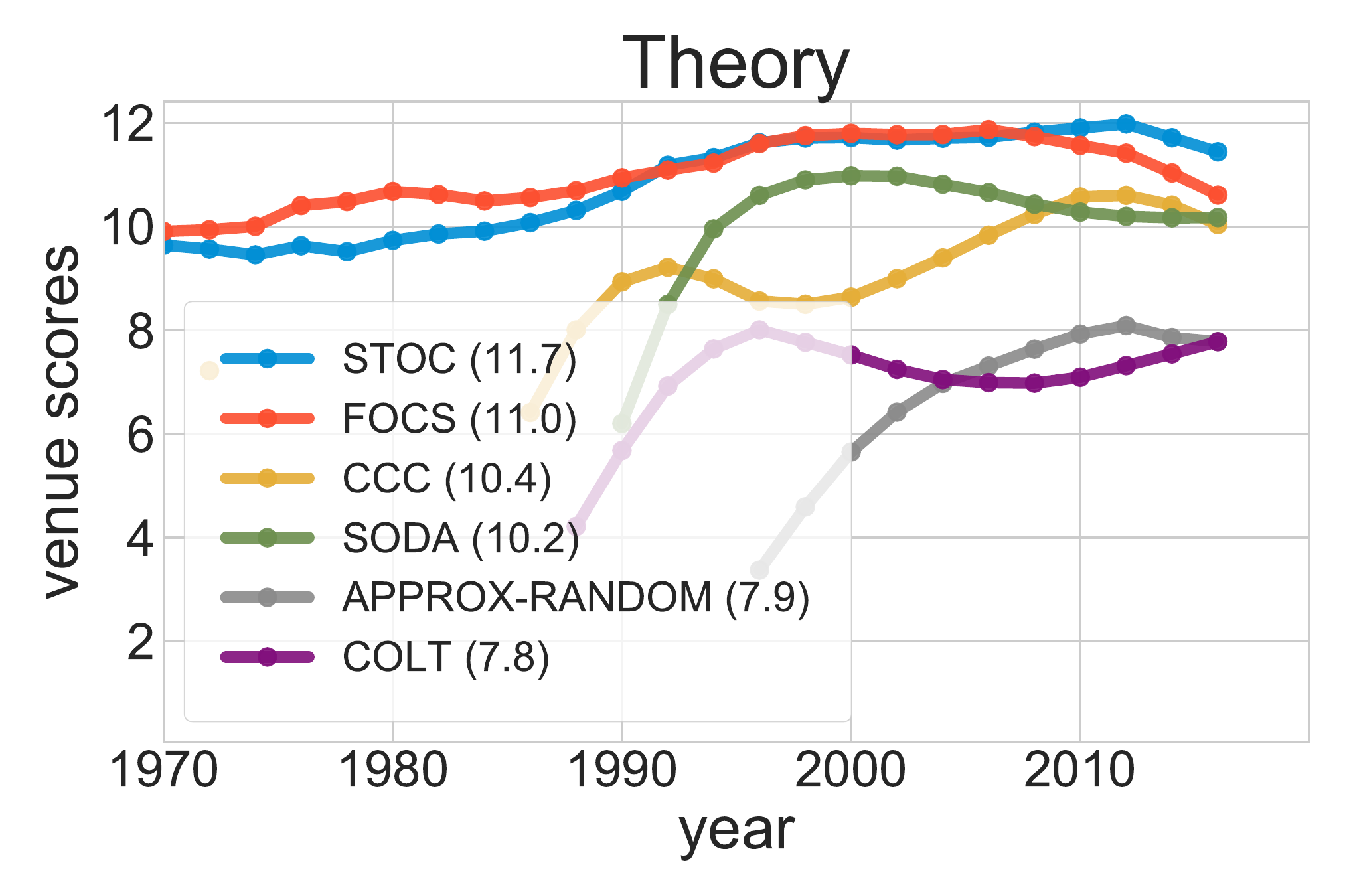}
    \includegraphics[width=0.195\textwidth]{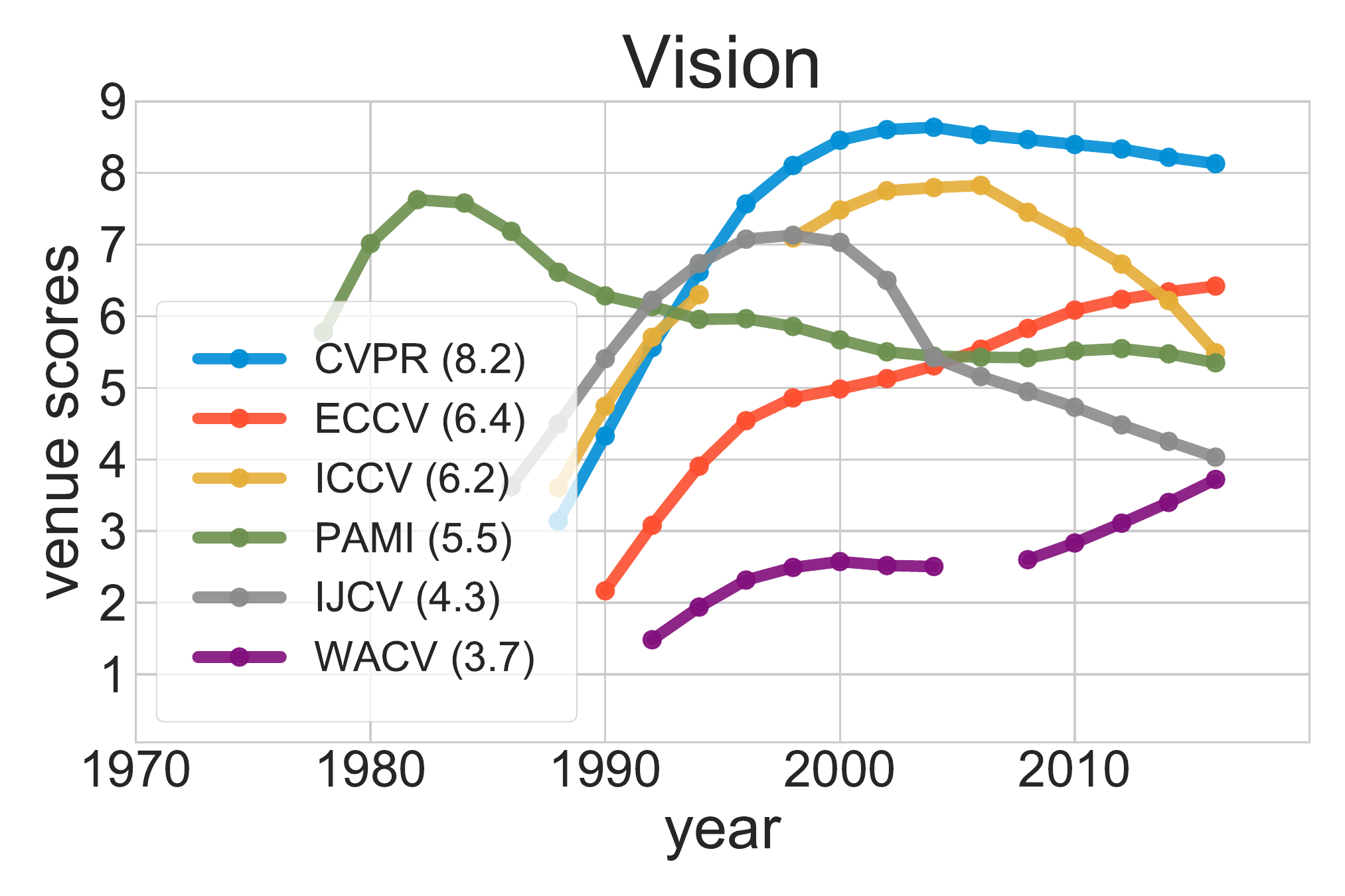}
        \includegraphics[width=0.195\textwidth]{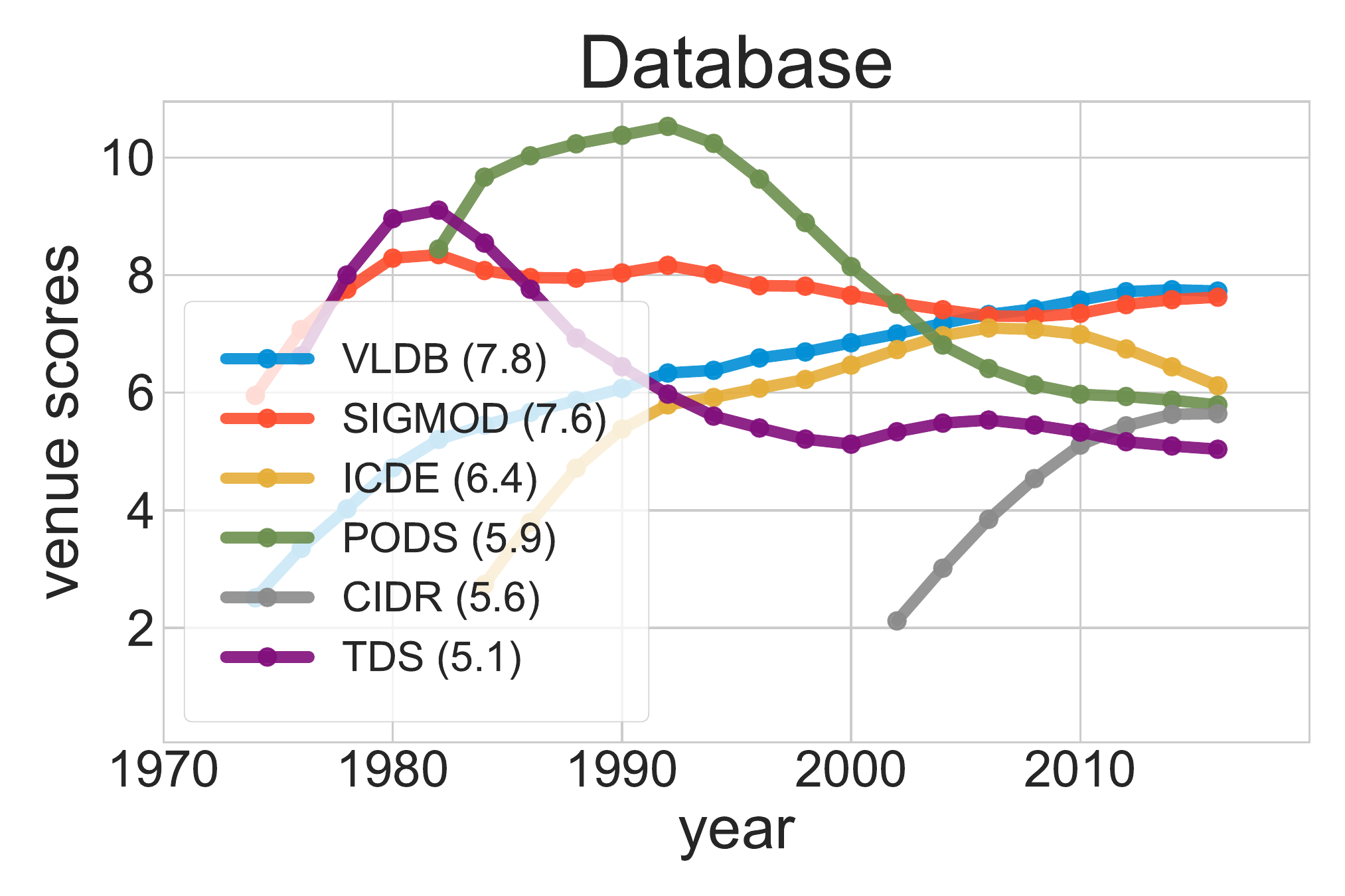}    \includegraphics[width=0.195\textwidth]{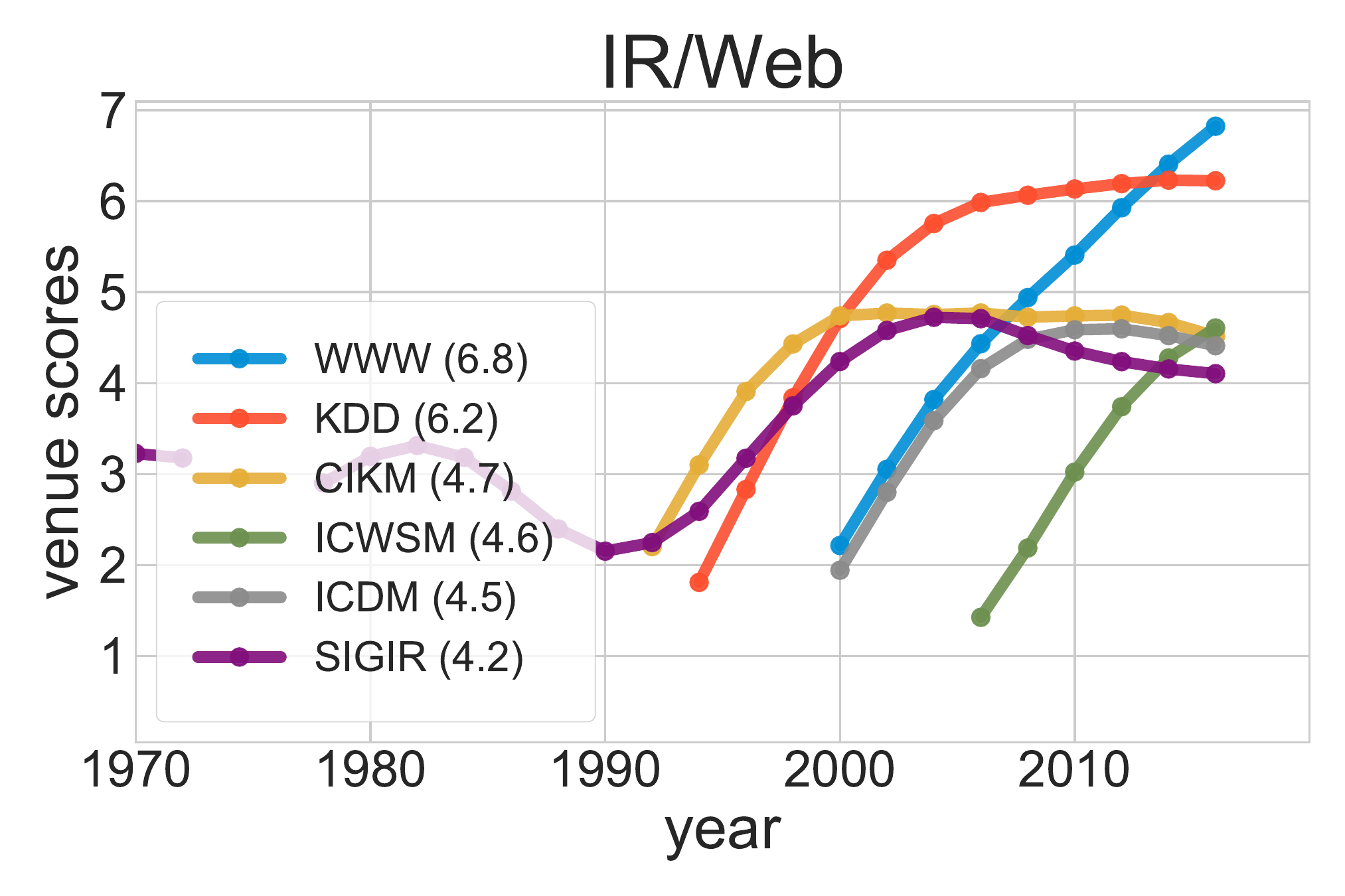}    \includegraphics[width=0.195\textwidth]{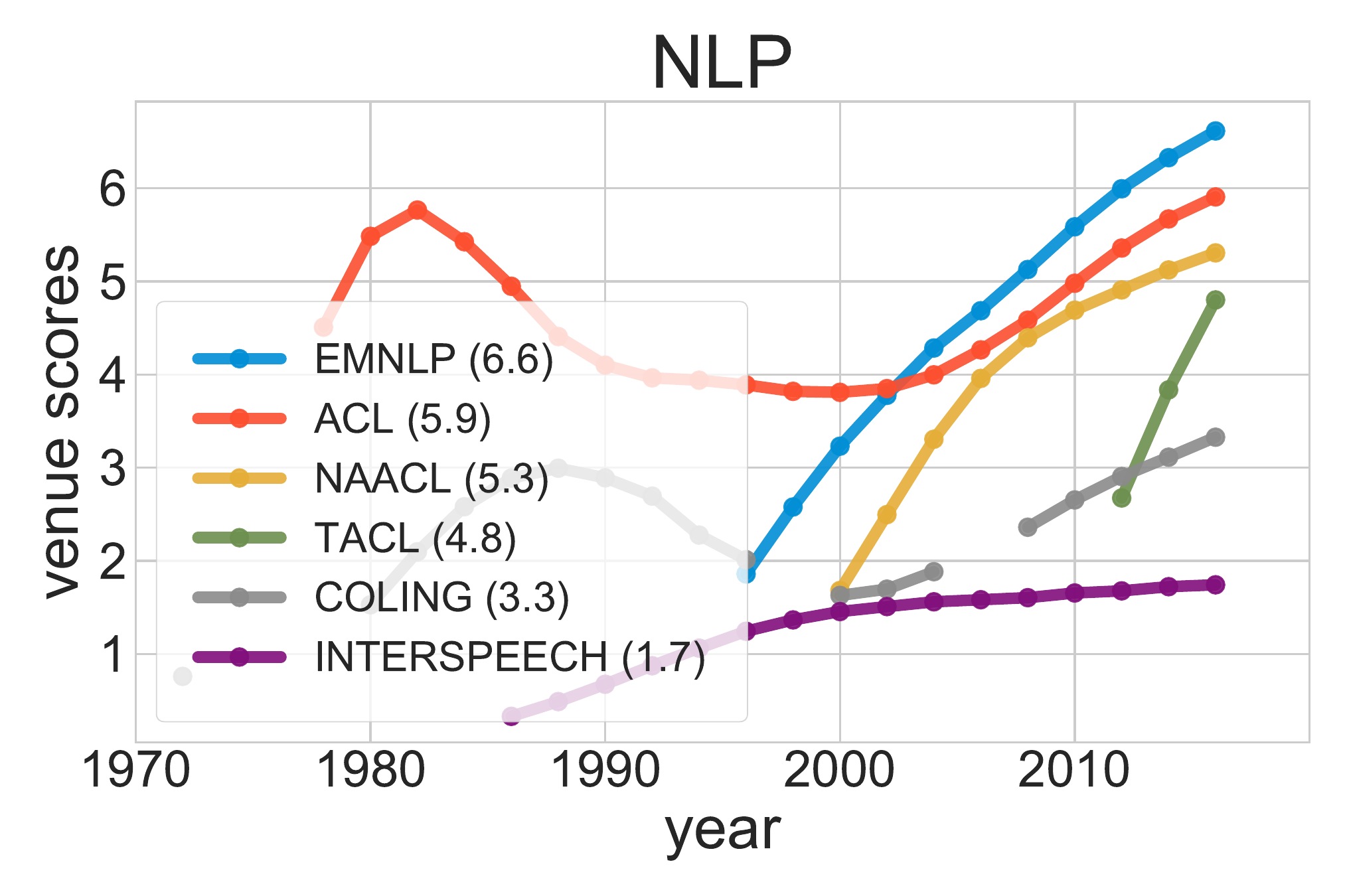}    \includegraphics[width=0.195\textwidth]{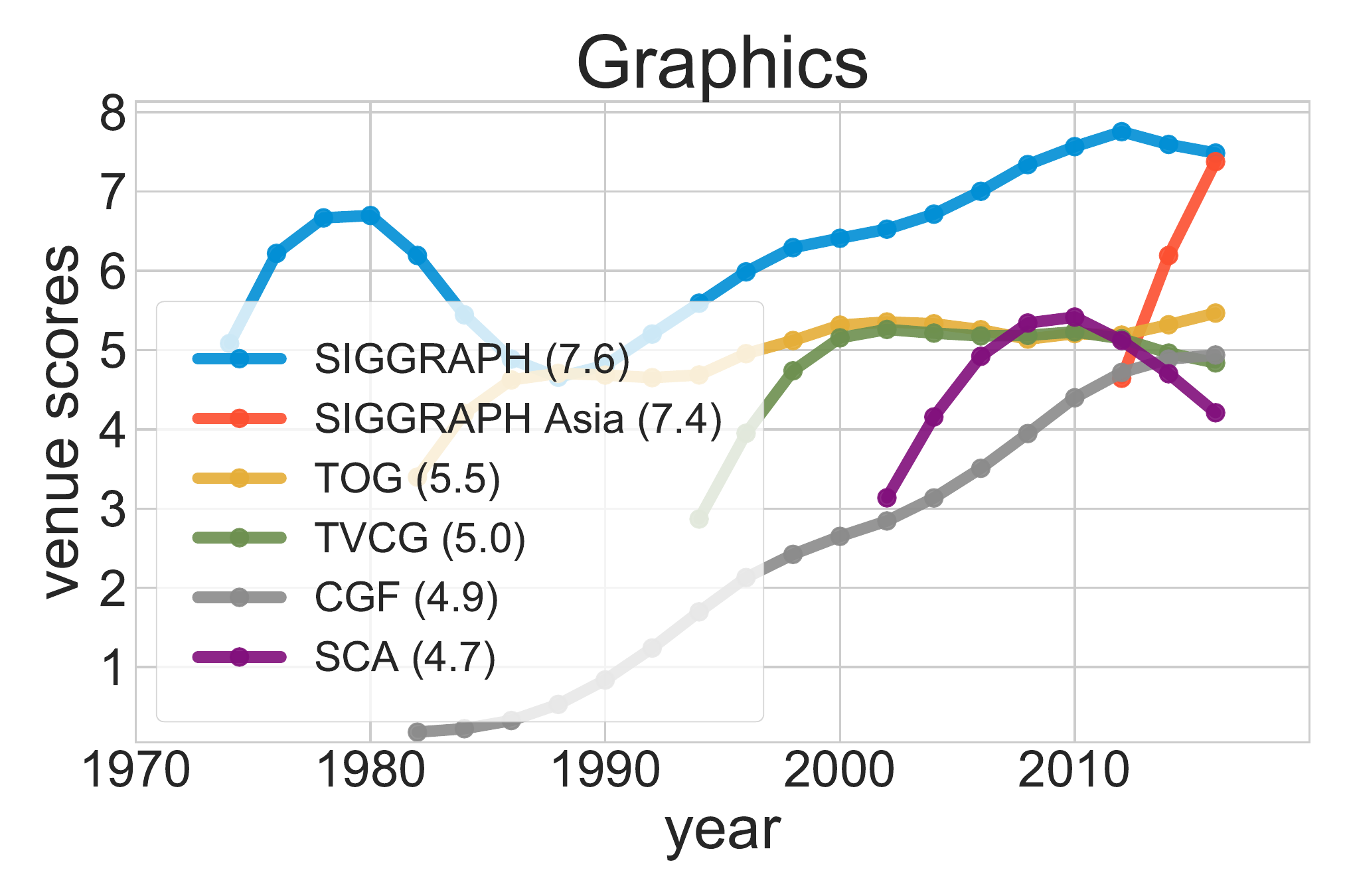}    \includegraphics[width=0.195\textwidth]{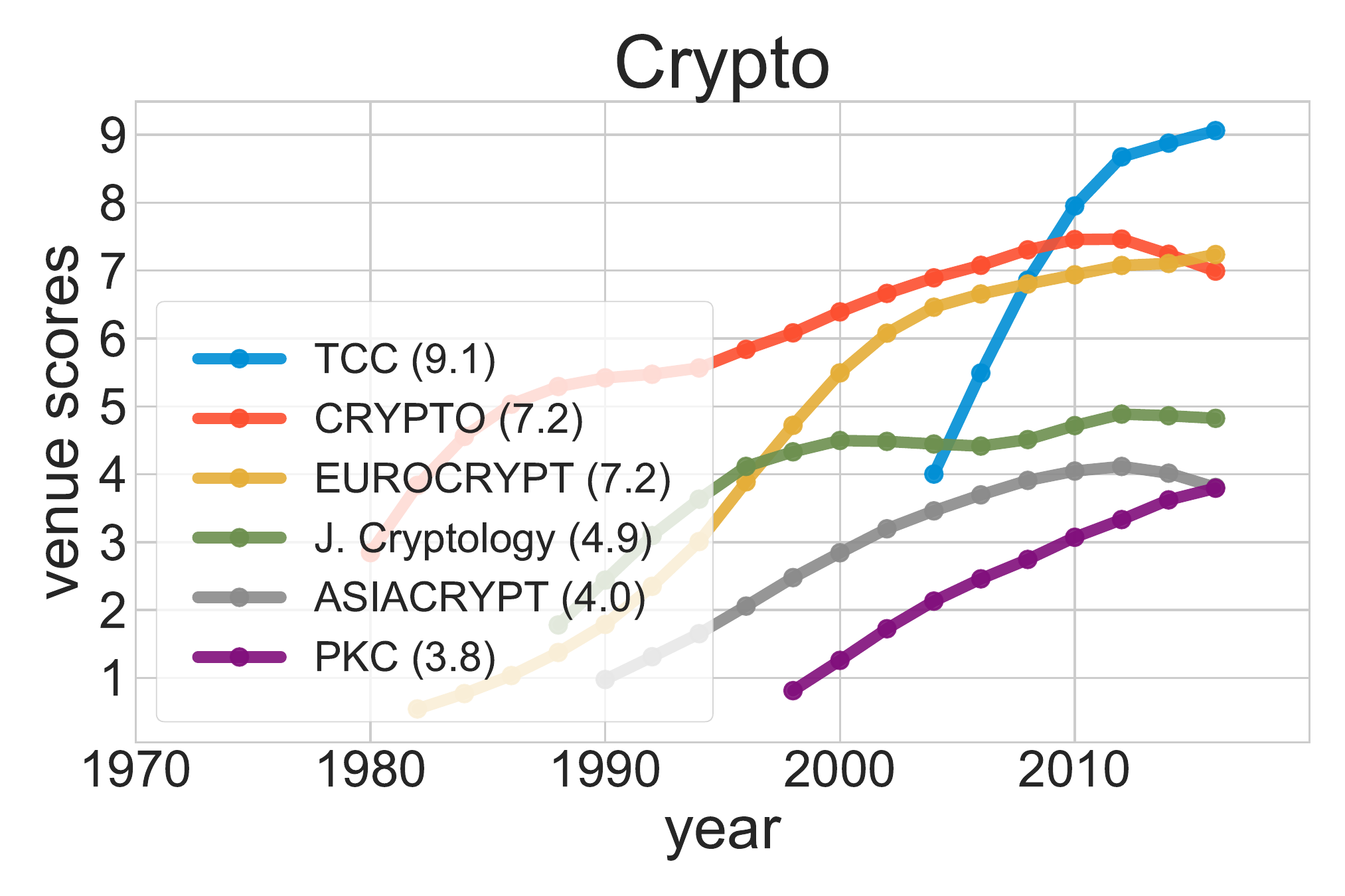}
  \caption{Results from our method, demonstrating differences and changes in conference quality over time in various subfields. These graphs includes size and year normalization and are the result of combining multiple different metrics of interest.}
  \Description{Correlation Results }
  \label{fig:teaser}
\end{teaserfigure}

%
% This command processes the author and affiliation and title information and builds
% the first part of the formatted document.
\maketitle

\section{Introduction}
There exist many tools to evaluate professional academic scholarship. For example Elseiver's Scorpus provides many author-level and journal-level metrics to measure the impact of scholars and their work~\cite{colledge2010sjr,da2017citescore}. Other publishers, such as the Public Library of Science, provide article-level metrics for their published work~\cite{fenner2013can}. Large technology companies, such as Google and Microsoft provide their own publicly available metrics for scholarship~\cite{pmid21814253}. Even independent research institutes, such as the Allen Institute's Semantic Scholar~\cite{ammar2018construction}, manage their own corpus and metrics for scholarly productivity. However, these author-based metrics (often derived from citation measurements) can be inconsistent, even across these large, established providers~\cite{da2018multiple}. 

In this work, we propose a method for evaluating a comprehensive collection of published academic work by using an external evaluation metric. By taking a large collection of papers and using only information about their publication venue, who wrote them, and when, we provide a largely automated way of discovering not only venue's value. Further, we also develop a system for  automatic organization of venues. This is motivated by the desire for an open, reproducible, and objective metric that is not subject to some of the challenges inherent to citation-based methods~\cite{da2018multiple,bornmann2008citation,galiani2017life}.

We accomplish this by setting up a linear regression from a publication record to some metric of interest. We demonstrate three valid regression targets: status as a faculty member (a classification task), awarded grant amounts, and salaries. By using DBLP~\cite{ley2002dblp} as our source of publication data, NSF grants as our source of awards, University of California data for salaries, and CSRankings~\cite{CSRankings} for faculty affiliation status, we're able to formulate these as large data regression tasks, with design matrix dimensions on the order of a million in each dimension. However, since these matrices are sparse, regression weights can be obtained efficiently on a single laptop computer. Details of our method are explained in section~\ref{sec:method}. 

We call our results \textbf{venue scores} and validate their performance in the tasks of evaluating conferences, evaluating professors, and ranking universities. We show that our venue scores correlate highly with other influence metrics, such as h-index~\cite{Hirsch2005}, citations or highly-influential citations~\cite{Valenzuela2015IdentifyingMC}. Additionally, we show that university rankings, derived from publication records correlate highly with both established rankings~\cite{USN18,Times18,QSRankings} and with recently published quantitative metrics~\cite{CSRankings,CSMetrics,ScholarRank,clauset2015systematic}. 

\section{Related Work}\label{sec:related}
Quantitative measures of academic productivity tend to focus on methods derived from citation counts.  By using citation count as the primary method of scoring a paper, one can decouple an individual article from the authors who wrote it and the venue it was published in. Then, robust citation count statistics, such as h-index~\cite{Hirsch2005}, can be used as a method of scoring either individual authors, or a specific venue. Specific critiques of h-index scores arose almost as soon as the h-index was published, ranging from a claimed lack-of-utility~\cite{lehmann2006measures}, to a loss of discriminatory power~\cite{Tol2008ARS}.

Citations can also be automatically analyzed for whether or not they're highly influential to the citing paper, producing a "influential citations" metric used by Semantic Scholar~\cite{Valenzuela2015IdentifyingMC}. Even further, techniques from graph and network analysis can be used to understand systematic relationships in the citation graph~\cite{bergstrom2007eigenfactor}. Citation-based metrics can even be used to provide a ranking of different universities~\cite{CSMetrics}. 

Citations-based metrics, despite their wide deployment in the scientometrics community, have several problems. For one, citation behavior varies widely by fields~\cite{bornmann2008citation}. Additionally, citations often exhibits non-trivial temporal behavior~\cite{galiani2017life}, which also varies greatly by sub-field. These issues highly affect one's ability to compare across disciplines and produce different scores at different times. Recent work suggests that citation-based metrics struggle to effectively capture a venue's quality with a single number~\cite{Walters17}. Comparing citation counts with statistical significance requires an order-of-magnitude difference in the citation counts~\cite{Kurtz17}, which limits their utility in making fine-grained choices. Despite these quality issues, recent work~\cite{ScholarRank} has demonstrated that citation-based metrics can be used to build a university ranking that correlates highly with peer assessment; we show that our method provides a similar quality of correlation. 

Our use of straightforward publication data (Section~\ref{sec:data}) enables a much simpler model. This simplicity is key, as the challenges in maintaining good citation data have resulted in the major sources of h-index scores being inconsistent with one another~\cite{da2018multiple}. 

While there exist many forms of ranking journals such as Eigenfactor~\cite{bergstrom2007eigenfactor} or SJR~\cite{doi:10.1096/fj.08-107938}, these tend to focus on journal-level metrics while our work focuses on all venues, including conferences.

\subsection{Venue Metrics}
We are not the first to propose that scholars and institutions can be ranked by assigning scores to published papers\cite{ren2007automatic}. However, in prior work the list of venues is often manually curated and assigned equal credit, a trend that is true for studies in 1990s~\cite{Geist:1996:CRP:240483.240505} and their modern online versions~\cite{CSRankings}. Instead, we propose a method for obtaining automatic scores for each venue, and in doing so, requires no manual curation of valid venues. 

Previous work~\cite{yan2007toward} has developed methods for ranking venues automatically, generating unique scores based only on author data by labeling and propagating notions of "good" papers and authoritative authors. However, this work required a manually curated seed of what good work is. It was only demonstrated to work on a small sub-field of conferences, as new publication cliques would require new labeling of "good" papers. Recent developments in network-based techniques for ranking venues have included citation information~\cite{zhang2018ranking}, and are able to produce temporal models of quality. In contrast, our proposed model doesn't require citation data to produce sensible venue scores. 

%Some people rank venues, other people prioritize venues~\cite{neumann2018prioritizing}.  University rankings can be prone to manipulation~\cite{Johnes2018}. 

Our work, in some ways, is most similar to that of CSRankings~\cite{CSRankings}. 
CSRankings maintains a highly curated set of top-tier venues; venues selected for inclusion are given 1 point per paper, while excluded venues are given 0 points. Additionally, university rankings produced by CSRankings include a manually curated set of categories, and rankings are produced via a geometric mean over these categories. In comparison, in this work, we produce unique scores for every venue and simply sum together scores for evaluating authors and institutions. 

In one of the formulations of our method, we use authors status as faculty (or not faculty) to generate our venue scores. We are not alone in this line of analysis, as recent work has demonstrated that faculty hiring information can be used to generate university prestige rankings~\cite{clauset2015systematic}. 

Many existing approaches either focus only on journals~\cite{doi:10.1096/fj.08-107938}, or do not have their rankings available online. For our dataset, we do not have citation-level data available, so we are unable to compare against certain existing methods on our dataset. However, as these methods often deploy a variant of PageRank~\cite{page1999pagerank}, we describe a PageRank baseline in Section~\ref{sec:pagerank} and report its results.

\section{Data} \label{sec:data}
Our primary data is the \textit{dblp computer science bibliography}~\cite{ley2002dblp}. DBLP contains millions of articles, with millions of authors across thousands of venues in Computer Science. We produced the results in this paper by using the dblp-2019-01-01 snapshot. We restricted ourselves to only consider conference and journal publications, skipping books preprints, articles below 6 pages and over 100 pages. We also merged dblp entries corresponding to the same conference. This lead to a dataset featuring 2,965,464 papers, written by 1,766,675 authors, across 11,255 uniquely named venues and 50 years of publications (1970 through 2019). 

Our first metric of interest is an individual's status as a faculty member at a university. For this, we used the faculty affiliation data from  CSRankings~\cite{CSRankings}, which are manually curated and contain hundreds of universities across the world and about 15,000 professors. For evaluation against other university rankings, we used the ScholarRank~\cite{ScholarRank} data to obtain faculty affiliation, which contains a more complete survey of American universities (including more than 50 not currently included in CS Rankings). While CSRankings data is curated to have correct DBLP names for faculty, the ScholarRank data does not. To obtain a valid affiliation, the names were automatically aligned with fuzzy string matching, resulting in about 4,000 faculty with good seemingly unique DBLP names and correct university affiliation. Although those two methods are manually curated, automatic surveys of faculty affiliations have recently been demonstrated~\cite{automaticsurvey2018}. 

Our second metric of interest was National Science Foundation grants, where we used awards from 1970 until 2018. This data is available directly from the NSF~\cite{NSFAwards}. We adjusted award amounts using annual CPI inflation data. We restricted ourselves to awards that had finite amount, where we could match at least half the Principal Investigators on the grant to DBLP names and the grant was above \$20,000. Award amounts over 10 million dollars were clipped in a smooth way to avoid matching to a few extreme outliers. This resulted in 407,012 NSF grants used in building our model. 

Our third and final metric of interest was University of California salary data~\cite{UCSalaries}. This was inspired by a paper that predicted ACM/IEEE Fellowships for 87 professors and used salary data~\cite{academicMoneyball}. We looked at professors across the entire University of California system, matching their names to DBLP entries in an automated way. We used the maximum salary amount for a given individual across the 2015, 2016 and 2017 datasets, skipping individuals making less than $120,000$ or over $800,000$ dollars. This resulted in 2,436 individuals, down from 3,102 names that we matched and an initial set of about 20,000 initial professors. As DBLP contains some Chemistry, Biology, and Economics venues, we expect that some of these are likely not Computer Science professors. 
\begin{table}[h]
\caption{Spearman's $\rho$ correlation between rankings produced by targeting different metrics of interest.}
\label{tab:corrtable}
\begin{tabular}{lrrr}
\toprule
{} &   Faculty &       NSF &    Salary \\
\midrule
Faculty &  1.00 &  \textbf{0.91} &  0.84 \\
NSF     &  \textbf{0.91} &  1.00 &  \textbf{0.86} \\
Salary  &  0.84 &  0.86 &  1.00 \\
\bottomrule
\end{tabular}
\end{table}
\section{Method}\label{sec:method}
Our basic model is that a paper in a given publication venue (either a conference or a journal), having passed peer review, is worth a certain amount of value. Certain venues are more prestigious, impactful or selective, and thus should have higher scores. Other venues have less strict standards, or perhaps provide less opportunity to disseminate their ideas~\cite{Morgan2018} and should be worth less. While this model explicitly ignores the differences in paper quality at a publication venue, discarding this information enables the use of a large quantity of data to develop a statistical scoring system. 

This methodology is not valid for all fields of science, nor all models of how impactful ideas are developed and disseminated. Our method requires individual authors have multiple publications across many different venues, which is more true in Computer Science than the natural sciences or humanities, where publishing rates are lower~\cite{publishrates18}. If instead we assume that all research ideas produce only a single paper, or that passing peer-review is a noisy measurement of quality~\cite{nips2016review}, then our proposed method would not work very well. Instead, the underlying process which supports our methodology is that good research ideas produce multiple research publications in selective venues; better ideas would produce more individual publications in higher quality venues. The concept of \textit{All models are wrong, but some are useful} is our guide here. This assumption allows us to obtain venue scores in an automatic way, and then use these scores to evaluate both authors and institutions. 

Our approach to ranking venues is to construct a regression task, apply an optimization method to solve it, and use the resulting weights. Optimization's ability to generate powerful representations is well-studied in machine learning~\cite{Rumelhart1986} and statistics~\cite{projectionpursuit}. The resulting weights are often useful~\cite{coates2011importance}, and this methodology is widely used in natural language processing~\cite{mikolov2013efficient,pennington2014glove}. 

\subsection{Formal Setup}
In general, we will obtain a score for each venue by setting up a linear regression task in the form of equation~\ref{eq:facultyregress}.
\begin{equation}\label{eq:facultyregress}
    \bordermatrix{& \text{conf}_1 & \text{conf}_2 &\ldots & \text{conf}_n & \cr
                    \text{auth}_1 & 1 &  3  & \ldots & 0 &1 \cr
                    \text{auth}_2& 1  &  0 & \ldots & 1 &1\cr
                    \vdots & \vdots & \vdots & \ddots & \vdots & \vdots \cr
                    \text{auth}_m& 0  &   2       &\ldots & 4 & 1} 
    \begin{pmatrix}
    x_0 \\ x_1 \\ \vdots \\ x_n
    \end{pmatrix} = \begin{pmatrix}
    \text{isProf}_1 \\ \text{isProf}_2 \\ \vdots \\ \text{isProf}_m
    \end{pmatrix} 
\end{equation}
Authors are listed along the rows, and venues are listed along the columns. The number of publications that an author has in a conference is noted in the design matrix. There is an additional column of 1s to learn a bias offset in the regression. 

Different forms of counting author credit are discussed in section~\ref{sec:authormodel}, while different regression targets are discussed in section~\ref{sec:data}. In equation~\ref{eq:facultyregress}, the regression target is shown as a binary variable indicating whether or not that author is currently a professor. If this linear system, $Ax = b$ is solved, then the vector $x$ will contain real-valued scores for every single publishing venue. Since our system is over-determined, there is generally no exact solution. 

Instead of solving this sparse linear system directly, we instead solve a regularized regression, using a robust loss and $L_2$ regularization. That is, we iteratively minimize the following expression via stochastic gradient descent~\cite{robbins1951stochastic}
\begin{equation}\label{eq:loss}
L(Ax,b) + \lambda ||x||^2
\end{equation}
The $L_2$ regularization enforces a Gaussian prior on the learned conference scores. We can perform this minimization in Python using common machine learning software packages~\cite{scikit-learn}. We tend to use a robust loss function, such as the \textit{Huber loss} in the case of regression~\cite{huber1964}, which is quadratic for errors of less than $\delta$ and linear for errors of larger than $\delta$. It can be written as
\begin{equation}\label{eq:huber}
    L(\hat{y},y)= 
\begin{cases}
    \frac{1}{2} (y-\hat{y})^2,& \text{if } |y-\hat{y}| \leq \delta \\
    \delta |y-\hat{y}| - \frac{1}{2}\delta^2,              & \text{otherwise}
\end{cases}
\end{equation}
In the case of classification, we have labels $y \in \{-1,1\}$ and use the \textit{modified Huber loss}~\cite{modifiedhuber}, 
\begin{equation} \label{eq:modifiedhuber}
    L(\hat{y},y)= 
\begin{cases}
    \max(0,1-y\hat{y})^2,& \text{if } y\hat{y} \geq -1 \\
     -4y\hat{y},              & \text{otherwise}
\end{cases}
\end{equation}

We experimented with other loss functions, such as the logistic loss, and while they tended to produce similar rankings and results, we found that the modified Huber loss provided better empirical performance in our test metrics, even though the resulting curves looked very similar upon qualitative inspection. 
\clearpage
\subsection{Metrics of Interest}
As detailed in section~\ref{sec:data}, we targeted three metrics of interest: status as a faculty member, NSF award sizes, and professor salaries. Each of these metrics came from a completely independent data source, and we found that they each had their own biases and strengths (more in section \ref{sec:results}). 

For faculty status classification, we used the \textit{modified huber loss} and CSRankings~\cite{CSRankings} faculty affiliations. To build venue scores that reward top-tier conferences more highly, we only gave professors in the top-$k$ ranked universities positive labels. We tried $k=5,16,40,80$, and found that we got qualitatively different results with quantitative similar performance. Unless otherwise stated, we used $k=40$. The university ranking used to select top-$k$ was CSRankings itself, and included international universities, covering the Americas, Europe and Asia.  This classification is performed across all authors, leading to 1.7 million rows in our design matrix. 

For the NSF awards, every Principal Investigator on the award had their papers up to the award year as the input features. We used a Huber loss, $\lambda=0.03$, and experimented with normalizing our award data to have zero mean and unit variance. Additionally, we built models for both raw award sizes and log of award sizes; the raw award sizes seem to follow a power-law while the log award sizes seem distributed as approximately a Gaussian. Another model choice is whether to regress each NSF grant as an independent measurement or instead a marginal measurement which tracks the cumulative total of NSF grants received by the authors. If not all authors on the grant were matched to DBLP names, we only used the fraction of the award corresponding to the fraction of identified authors. This regression had $\sim \frac{1}{2}$ million rows in its design matrix . 

For the salary data, we found that normalizing the salary data to have zero mean and unit variance led to a very poor regression result, while having no normalization produced a good result. This regression only had $\sim 2,400$ datapoints, and thus provided information about fewer venues than the other metrics of interest. 

\subsection{Modeling Change Over Time}\label{sec:temporal}

In modeling conference values, we wanted to build a model that could adjust for different values in different years. For example, a venue may be considered excellent in the 1980s, but may have declined in influence and prestige since then. To account for this behavior, we break our dataset into chunks of $n$ years and create a different regression variable for each conference for each chunk. The non-temporal model is obtained simply by setting $n \geq 50$.

\begin{figure}
  \centering
  \includegraphics[width=\linewidth]{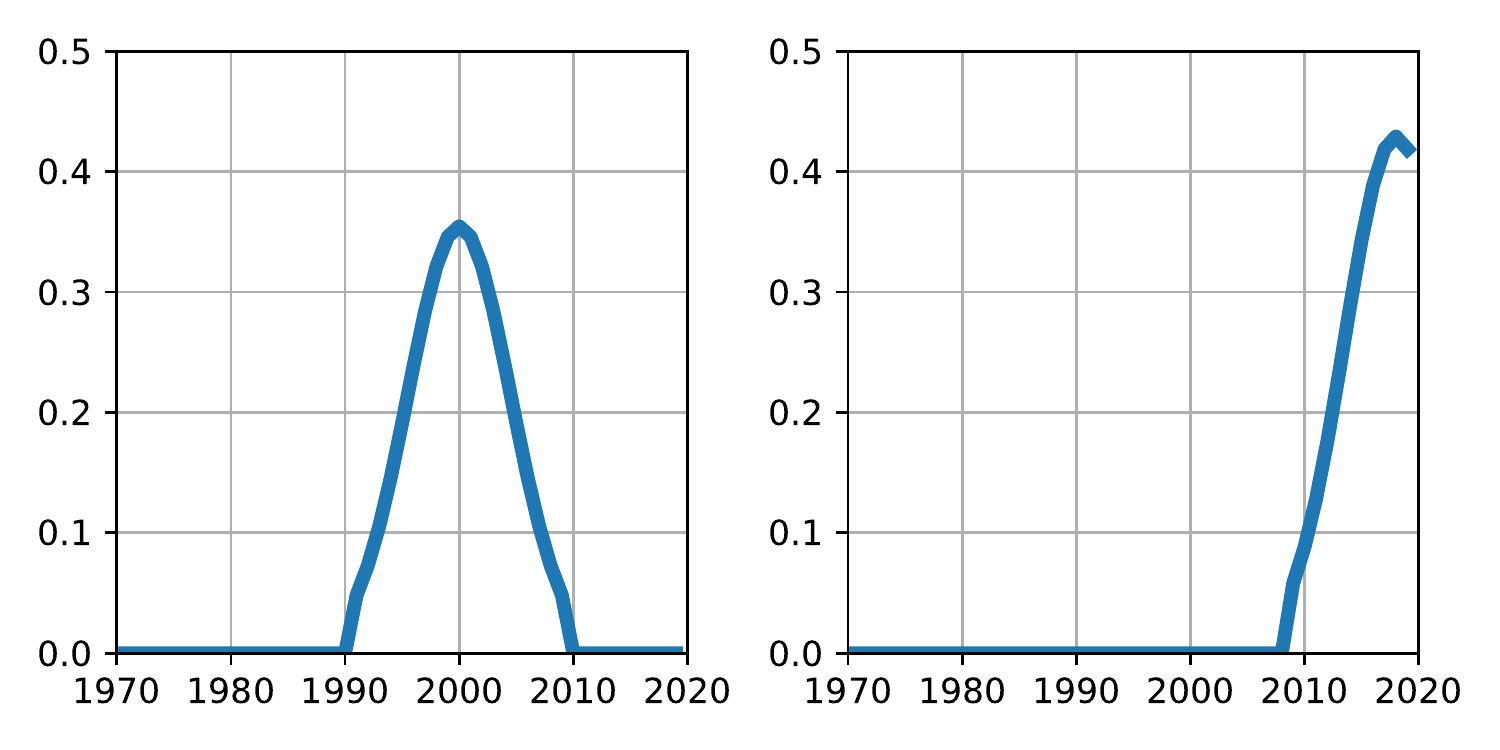}
  \caption{Truncated Gaussian ($\sigma=4.5$) used to splat a publication's value across multiple years. Examples centered at the year 2000 and the year 2018}
  \Description{Truncated Gaussian}
  \label{fig:tgauss}
\end{figure}

We also examine a model that creates an independent variable for each year, for each conference. After setting $n=1$ in the block model, we splat each publication as a Gaussian at a given year. By modifying $\sigma$, we can control the smoothness of the obtained weights. The Gaussian is applied via a sparse matrix multiply of the design matrix $A$ with the appropriate band diagonal sparse matrix $G$. The use of a truncated Gaussian (where $p < 0.05$ is clipped and the Gaussian is re-normalized) enables our matrix to maintain sparsity. We used a $\sigma=4.5$, which produced an effective window size of about 10 years. This can be seen visually in Figure~\ref{fig:tgauss}. 

Our different temporal models are compared in Table~\ref{tab:temporalresults} by correlating against existing author-level, journal-level and university-level metrics. For evaluation details see Section~\ref{sec:eval}. 

\subsection{Normalizing Differences Across Years}\label{sec:norm_year}

\begin{figure}
  \centering
   \includegraphics[width=0.99\linewidth]{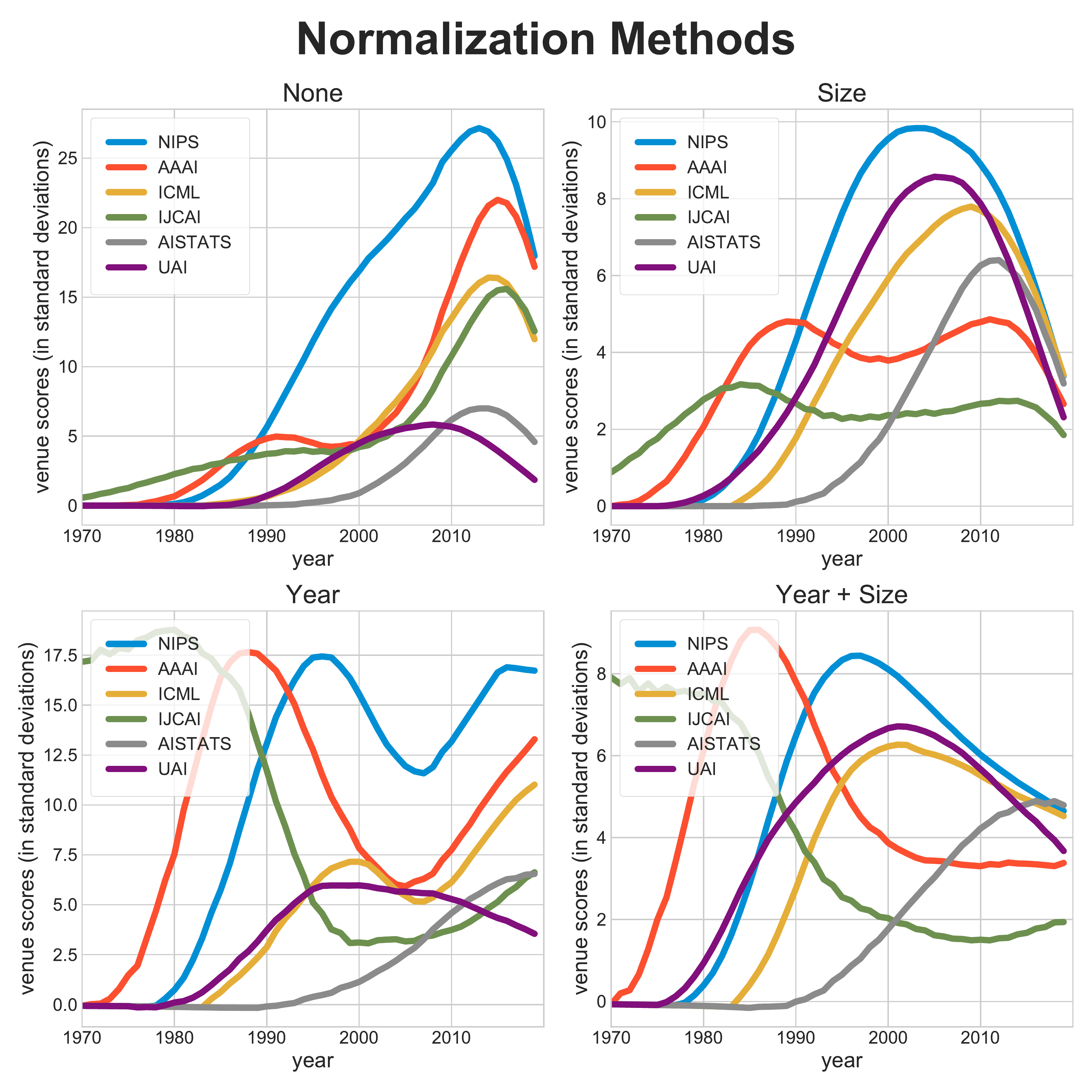}
  \caption{Results showing the effect of performing a normalization for venue year and size. See Sections \ref{sec:norm_year} and \ref{sec:confsize}.  }
  \label{fig:adjust}
\end{figure}

The temporal models described in the previous section have an inherent bias. Due to the temporal window of the DBLP publication history, there is variation in distributed value due to changes annual NSF funding, the survivorship of current academic faculty, etc. To adjust for this bias, we scale each conference-year value by the standard deviation of conference values in that year. This scaling can help or hurt performance, depending on which metric of interest the model is built against. It generally produces flatter value scores over time but leads to some artifacts. The effects of this normalization are shown in Figure~\ref{fig:adjust} and Table~\ref{tab:normalizations}. In our final version, we instead normalize by the average of the top 10 conferences in each year, which produced flatter results over time, 

\clearpage
\subsection{Normalizing Differences In Venue Size}\label{sec:confsize}
Our model uses $L_2$ regularization on the venue scores, which tends to squash the value of variables with less explanatory power.This process often resulted in the under-valuing of smaller, more selective venues. To correct for this size bias, instead of giving each paper $1$ point of value in the design matrix, we give each paper $\frac{1}{M}$ credit, where $M$ is the number of papers at that venue in that year; this step is performed before Gaussian splatting. This produced rankings which emphasized small venues and provided a good notion of efficient venues. In practice, we instead used a blended result with a multiplier of $\frac{1}{\sqrt[\alpha]{M}}$ with $\alpha=1.5849$, the Hausdorff dimension of the Sierpinski triangle, an arbitrary constant.

\begin{table}[]
\centering
\caption{Spearman correlation between our model and existing metrics, showing the effect of different normalization schemes. See Sections \ref{sec:norm_year} and \ref{sec:confsize} for model details. See Section~\ref{sec:eval} for evaluation details. }
\label{tab:normalizations}
\begin{tabular}{@{}lp{1.2cm}p{1.2cm}p{1.5cm}p{1cm}@{}}
\toprule
       Normalization     & influential citations (author) & h-index (author) & h-index (university) & h-index (venue) \\ \midrule
None        & 0.72                           & 0.61             & 0.60                 & \textbf{0.26}            \\
Year        & 0.72                           & \textbf{0.66}             & 0.58                 & 0.18            \\
Size        & \textbf{0.74}                           & 0.63             & \textbf{0.61}                 & 0.25            \\
Year + Size & 0.72                           & 0.61             & 0.60                 & \textbf{0.26}            \\ \bottomrule
\end{tabular}
\end{table}

\subsection{Modeling Author Position} \label{sec:authormodel}
Another question to consider is how credit for a paper is divided up amongst the authors. We consider four models of authorship credit assignment:
\begin{enumerate}
    \item Authors get $\frac{1}{n}$ credit for each paper, where $n$ is the number of authors on the paper. Used by ~\cite{CSRankings}.
    \item All authors get full credit (1 point) for each paper
    \item Authors receive less credit for later positions ($\frac{1}{1},\frac{1}{2},\frac{1}{3}\cdots \frac{1}{n}$), normalized so total credit sums to 1). This model assigns more credit to earlier authors and is used by certain practitioners~\cite{Sekercioglu371,BestPapers}. 
    \item The same as (3), except the last author is explicitly assigned equal credit with the first author before normalization. 
\end{enumerate}
Using Spearman correlation with Semantic Scholar's "highly influential citations", an evaluation metric described in depth in Sec.~\ref{sec:authoreval}, we can evaluate each of these models. Specifically, there are two places where venues scores require a selection of authorship model. The first is how much credit is assigned to each paper when performing regression (in the case of our \textit{faculty metric of interest}). The second is when evaluating authors with the obtained regression vector. See table~\ref{tab:authormodels} for a summary of experimental results. 
\begin{table}
\caption{Correlation (Spearman's $\rho$) between our model and Semantic Scholar~\cite{Valenzuela2015IdentifyingMC}, showing the properties of different authorship models. For details see section~\ref{sec:authormodel}.}
\label{tab:authormodels}
\def\arraystretch{1.2}
\begin{tabular}{llllll}
\toprule
 &  & \multicolumn{4}{l}{Evaluation Author Model} \\
 &  & 1 & 2 & 3 & 4 \\ \hline
\multirow{4}{*}{\rotatebox[origin=c]{90}{\parbox{1.3cm}{Regression\\ Author Model}}} & 1 & 0.70 & \textbf{0.72} & 0.65 & 0.70 \\
 & 2 & 0.68 & \textbf{0.71} & 0.61 & 0.67 \\
 & 3 & \textbf{0.71} & \textbf{0.73} & \textbf{0.66} & \textbf{0.71} \\
 & 4 & 0.70 & \textbf{0.72} & 0.65 & \textbf{0.71} \\ \hline
\end{tabular}
\end{table}
For the purposes of evaluation, assigning full credit to authors (model 2) produced the best results, while model 3 consistently produced the lowest quality correlations. On the other hand, for the purposes of performing the classification task, the roles are flipped. Assigning full credit (model 2) consistently produces the worst quality correlations while using model 3 produces the highest quality correlations.  

\subsection{Combining Models}\label{sec:combining}
Since the proposed metrics of interest (faculty status, NSF awards, salaries) were generated from different independent regression targets, with different sized design matrices, there may be value in combining them to produce a joint model. The value of ensemble models is well documented in both theory~\cite{FREUND1997119} and practice~\cite{netflixprize09}. In the absence of a preferred metric with which to cross-validate our model, we simply perform an unweighted average of our models to obtain a \textit{gold} model. To ensure that the weights are of similar scale, the conference scores are normalized to have zero mean and unit variance before combining them. Venue scores that are too large or too small are clipped at 12 standard deviations. For the temporal models, this normalization is performed on a per-year basis. While table~\ref{tab:combining} shows results for a simple combination, one could average together many models with different choices of hyperparameters, regression functions, datasets, filters to scrub the data, etc. 

\begin{table}
  \caption{Correlation between our model and traditional measures of scholarly output on the dataset of CMU faculty. For model details see Section~\ref{sec:combining}. For evaluation details see Section~\ref{sec:authoreval}.}
  \label{tab:combining}
  \begin{tabular}{lcccc}
    \toprule
    Model &citations&h-index~\cite{Hirsch2005}&influential citations~\cite{Valenzuela2015IdentifyingMC}\\
    \midrule
    Faculty  & 0.59& 0.68& 0.71 \\ 
    NSF & 0.63& 0.66 & 0.67 \\ 
    Salary & 0.36& 0.36 & 0.41 \\ 
    Combined & \textbf{0.69}& \textbf{0.77} & \textbf{0.75} \\ 
  \bottomrule
\end{tabular}
\end{table}

\section{Results}\label{sec:results}
A visual example of some of venue scores is shown in Figure~\ref{fig:teaser}. We kept the y-axis fixed across all the different Computer Science sub-disciplines to show how venue scores can be used to compare different fields in a unified metric. There are additional results in our qualitative demonstration of normalization methods, Figure~\ref{fig:adjust}. 

Due to the variation in rankings produced by one's choice of hyperparameters, and the large set of venues being evaluated, we do not have a canonical set of rankings that can be presented succinctly here. Instead, we will focus on quantitative evaluations of our results in the following section. 

% AI and AH are author-level metrics on a dataset of faculty members. USN is a university-level ranking. VH and VC are venue-level metrics.
\begin{table}
\caption{Correlation between rankings produced by our model against rankings produced by traditional scholarly metrics. Different rows correspond to different hyperparameters choices for our model. Each column is a corresponds to a traditional metric. AI = Author Highly Influential Citations, AH = Author H-index, USN = US News 2018, VH = Venue H-index, VC = Venue Citations. }
\label{tab:temporalresults}
\begin{tabular}{@{}llp{0.7cm}p{0.7cm}p{0.7cm}p{0.7cm}p{0.7cm}@{}}
\toprule
 Years & Metric & AI & AH & USN & VH & VC \\ \midrule
$\sigma=4.5$ & Faculty & 0.73 & \textbf{0.69} & 0.74 & 0.63 & 0.42 \\
10 &  & 0.67 & 0.57 & \textbf{0.76} & 0.57 & 0.35 \\
50 &  & \textbf{0.75} & 0.68 & \textbf{0.76} & 0.38 & 0.21 \\ \midrule
$\sigma=4.5$ & NSF & 0.64 & 0.62 & 0.62 & 0.61 & 0.59 \\
10 &  & 0.68 & 0.68 & 0.60 & 0.59 & 0.60 \\
50 &  & 0.67 & 0.65 & 0.63 & \textbf{0.64} & \textbf{0.67} \\ \midrule
$\sigma=4.5$ & Salary & 0.62 & 0.58 & 0.59 & 0.48 & 0.55 \\
10 &  & 0.65 & 0.62 & 0.57 & 0.45 & 0.55 \\
50 &  & 0.66 & 0.63 & 0.56 & 0.43 & 0.63 \\ \bottomrule
\end{tabular}
\end{table}

%Some clear trends, such as the decline and resurgence of Artificial Intelligence research are clearly seen. 
\section{Evaluation}\label{sec:eval}
To validate the venue scores obtained by our regression methods, our evaluation consists of correlating our results against existing rankings and metrics. We consider three classes of existing scholarly measurements to correlate against: those evaluating universities, authors, and venues. Each of these classes has different standard techniques, and a different evaluation dataset, so they will be described separately in Sections~\ref{sec:unieval},~\ref{sec:authoreval}, and~\ref{sec:venueeval}.

In the case of our proposed method, venue scores, we have a simple way to turn them from a journal-based to an author-based or institution-based metric. Venues are evaluated directly with the scores. Authors are evaluated as the dot product of venue scores and the publication vector of an author. Universities are evaluated as the dot product of venue scores and the total publication vector of all faculty affiliated with that university.

\subsection{PageRank Baseline}\label{sec:pagerank}
Many existing approaches build on the idea of eigenvalue centrality~\cite{bergstrom2007eigenfactor,yan2007toward,zhang2018ranking}. We implemented PageRank~\cite{page1999pagerank} using the power iteration method to compute a centrality measure to use for both author-level and venue-level metrics. Unlike most versions of PageRank, which use citation counts, we implement two variants based solely on co-authorship information.

Author-level PageRank (PageRankA) is computed on the 1.7M x 1.7M sized co-authorship graph, where an edge is added for every time two authors co-author a paper. We found that the authors with highest centrality measures are often common names with insufficient disambiguation information in DBLP. 

Journal-level PageRank (PageRankC) is computed on the 11,000 x 11,000 co-authorship graph, where an edge is added for every author who publishes in both venues. When ran on the unfiltered DBLP data, the highest scoring venue was arXiv, an expected result.

\subsection{University Ranks}\label{sec:unieval}
For this work, we produce university rankings simply as an evaluation method to demonstrate the quality and utility of our venue scoring system. The reader is cautioned that university ranking systems can tend to produce undesirable gaming behavior~\cite{Johnes2018}, and are prone to manipulation.  

We obtained and aligned many existing university rankings for Computer Science departments. These include rankings curated by journalistic sources, such as the US News Rankings~\cite{USN18}, the QS Rankings~\cite{QSRankings}, Shanghai Ranking~\cite{Shanghai15}, Times Higher Education Rankings~\cite{Times18} and the National Research Council report~\cite{clauset2015systematic}. In addition, we consider purely quantitative evaluation systems such as ScholarRank~\cite{ScholarRank}, CSRankings~\cite{CSRankings}, CSMetrics~\cite{CSMetrics}, and Prestige Rankings~\cite{clauset2015systematic}. We additionally include ScholarRank's \textit{t10sum} across the matched faculty that our venue scores result uses. 

We follow a recent paper~\cite{ScholarRank}, which demonstrated the efficacy of a citation-based metric in producing rankings with large correlation against US News rankings. We extend these experiments to include more baselines. In contrast with ~\cite{ScholarRank}, we use a rank correlation metric (namely Kendall's $\tau$), which naturally handles ordinal ranking systems. While ScholarRank~\cite{ScholarRank} claimed a correlation of $> 0.9$ with US News, this was under Pearson's correlation coefficient, and the result under Kendall's $\tau$ is 0.768 in the published version and 0.757 using full precision ScholarRank. 

Our faculty-based regression is able to generate a result with the highest correlation against the US News rankings. We perform even better than ScholarRank, which was designed to optimize this metric (although under a non-rank correlation metric). 

% Please add the following required packages to your document preamble:
% \usepackage{booktabs}
\begin{table}
\caption{Section~\ref{sec:unieval}. Kendall's $\tau$ correlation across different University Rankings.}
\label{tab:usnews2018}
\begin{tabular}{@{}lp{2.4cm} @{}}
\toprule
Ranking & Correlation with US News 2018 \\ \midrule
USN2018~\cite{USN18} & \qquad1.000 \\
USN2010~\cite{clauset2015systematic} & \qquad0.928 \\
\textbf{Venue Scores} & \qquad\textbf{0.780} \\
ScholarRank~\cite{ScholarRank} & \qquad0.768 \\
ScholarRankFull & \qquad0.757 \\
CSMetrics~\cite{CSMetrics} & \qquad0.746 \\
CSRankings~\cite{CSRankings} & \qquad0.724 \\
Times~\cite{Times18} & \qquad0.721 \\
NRC95~\cite{clauset2015systematic} & \qquad0.713 \\
t10Sum~\cite{ScholarRank} & \qquad0.713 \\
Prestige~\cite{clauset2015systematic} & \qquad0.666 \\
Citations~\cite{ScholarRank} & \qquad0.665 \\
Shanghai~\cite{Shanghai15} & \qquad0.586 \\
\# of papers & \qquad0.585 \\
BestPaper~\cite{BestPapers} & \qquad0.559 \\
PageRankA & \qquad0.535 \\
PageRankC & \qquad0.532 \\
QS~\cite{QSRankings} & \qquad0.518 \\ \bottomrule
\end{tabular}
\end{table}

\begin{table}
  \caption{ Spearman's $\rho$ correlation between different journal-level metrics (N=1008). For details see Section~\ref{sec:venueeval}.}
  \label{tab:venuecorr}
\begin{tabular}{@{}llllp{1.1cm}p{1cm}@{}}
\toprule
& papers        & citations & h-index    & PageRank C & venue scores       \\ \midrule
papers        &       & \textbf{0.75} & 0.41      & \textbf{0.91}          & 0.25 \\
citations     & 0.75      &  & 0.60      & 0.69          & 0.27 \\
h-index             & 0.41      & 0.60 &      & 0.37          & \textbf{0.65} \\
PageRankC     & \textbf{0.91}      & 0.69 & 0.37      &           & 0.27 \\
venue scores & 0.25      & 0.27 & \textbf{0.65}      & 0.27          &  \\ \bottomrule
\end{tabular}
\end{table}
\subsection{Journal-level metrics}\label{sec:venueeval}
To evaluate the fidelity of our venue scores for journals and conferences, we obtain the h-index~\cite{Hirsch2005} and citation count for 1,308 conferences from Microsoft Academic Graph~\cite{ConfRanks}. We continue to use Spearman's $\rho$ as our correlation metric, even though rank-correlation metrics can be highly impacted by noisy data~\cite{abdullah1990robust}.

Under this metric, \textit{venue scores} correlated highly with \textit{h-index}. Notably, h-index and venue scores are highly correlated with each other, even though venue scores are much less correlated with raw conference size. See Figure~\ref{tab:venuecorr} for detailed results.

\subsection{Author-level Metrics}\label{sec:authoreval}
To evaluate our venue scores in the application of generating author-level metrics, we will use rank correlation (also known as Spearman's $\rho$)~\cite{spearman1904proof} between our venue scores and traditional author-level metrics such as h-index. Google Scholar was used to obtain citations, h-index~\cite{Hirsch2005}, i10-index, and Semantic Scholar used to obtain highly influential citations~\cite{Valenzuela2015IdentifyingMC}. Prior work has critiqued the h-index measure~\cite{yong2014critique} and proposed an alternative metric, derived from a citation count. However, our use of a rank correlation~ means that monotonically transformed approximations of citation counts would lead to identical scores. 

For evaluation, we collected a dataset for the largest Computer Science department in CSRankings ($n=148$). The results are shown in Table~\ref{tab:authorlevel}. We can see that venue scores highly correlate with h-index, influential citations, i10-scores and CSRankings scores. The results from the author-based PageRank are surprisingly similar to our venue scores. However, the conference-based PageRank performed worse than venue scores on every correlation metric. 

 \begin{table*}
\caption{Correlation between different author-level metrics for a dataset of professors (N=148). Details are in Section~\ref{sec:authoreval}. }
\label{tab:authorlevel}
\begin{tabular}{@{}llllllllll@{}}
\toprule
& papers       & citations & h-index & i10  & CSR~\cite{CSRankings}  & venue score & PageRankA & PageRankC & influence~\cite{Valenzuela2015IdentifyingMC}      \\ \midrule
papers       &       & 0.66    & 0.79 & 0.81      & 0.71         & \textbf{0.94}      & \textbf{0.94}      & 0.89      & 0.76 \\
citations    & 0.66      &     & 0.93 & 0.88      & 0.49         & 0.66      & 0.68      & 0.60      & \textbf{0.81} \\
h-index      & 0.79      & \textbf{0.93}    &  & \textbf{0.97}      & 0.56         & 0.75      & 0.81      & 0.68      & 0.80 \\
i10          & 0.81      & 0.88    & \textbf{0.97} &       & 0.53         & 0.75      & 0.82      & 0.69      & 0.73 \\
 CSR~\cite{CSRankings}    & 0.71      & 0.49    & 0.56 & 0.53      &          & 0.84      & 0.64      & 0.80      & 0.64 \\
venue score & \textbf{0.94}      & 0.66    & 0.75 & 0.75      & \textbf{0.84}         &       & 0.86      & \textbf{0.92}      & 0.78 \\
PageRankA    & \textbf{0.94}      & 0.68    & 0.81 & 0.82      & 0.64         & 0.86      &       & 0.83      & 0.72 \\
PageRankC    & 0.89      & 0.60    & 0.68 & 0.69      & 0.80         & 0.92      & 0.83      &       & 0.67 \\
influence~\cite{Valenzuela2015IdentifyingMC}     & 0.76      & 0.81    & 0.80 & 0.73      & 0.64         & 0.78      & 0.72      & 0.67      &  \\ \bottomrule
\end{tabular}
\end{table*}
\begin{figure}
  \centering  
  \includegraphics[width=\linewidth]{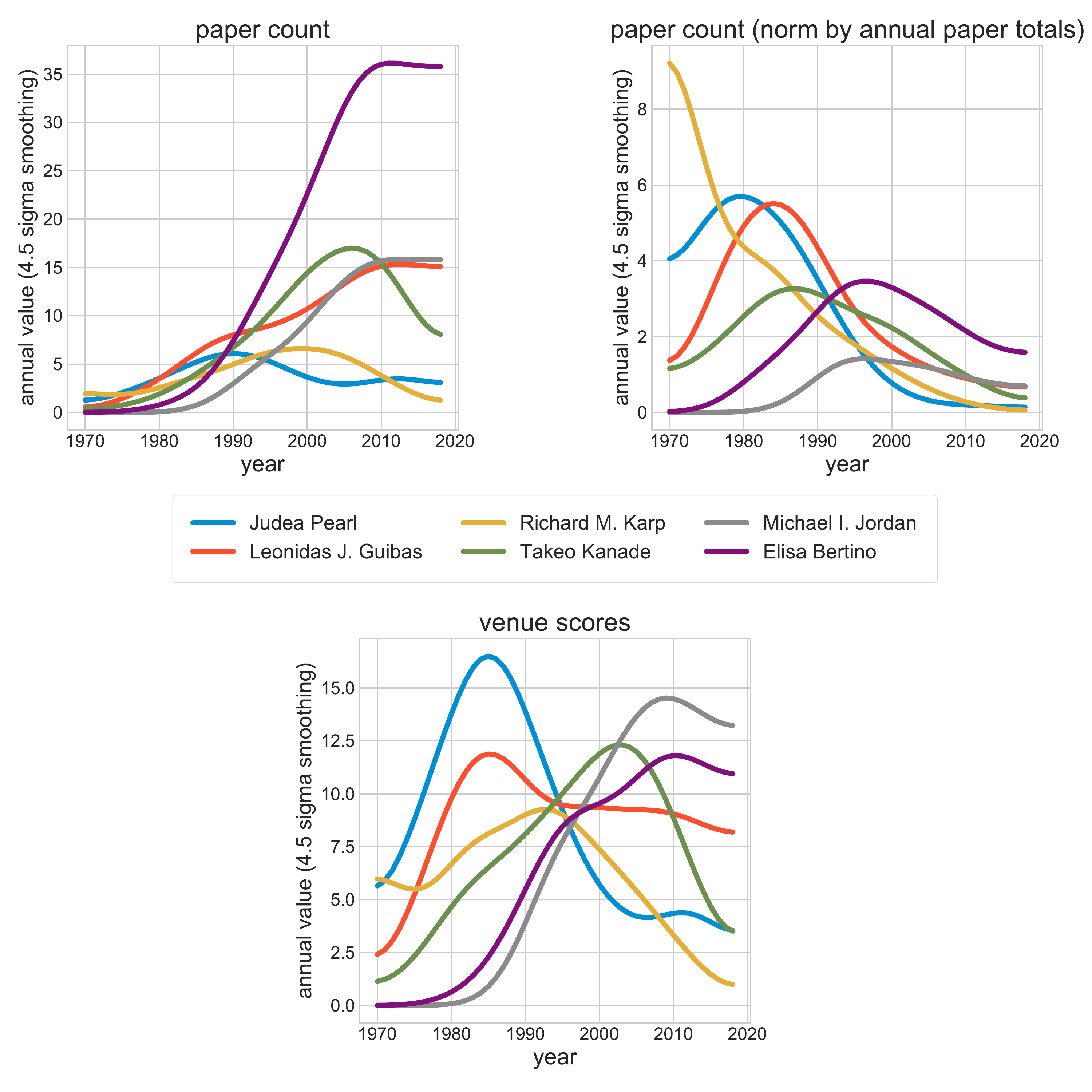}
  \caption{The career arcs of several accomplished Computer Scientists. The first row uses a simple model where all papers are all given equal weight; first using raw counts and then normalizing by the number of papers published each year. The second row shows our model.}
  \label{fig:fame}
\end{figure}
\section{Discussion}
Our results show a medium to strong correlation of venue scores against existing scholarly metrics, such as citation count and h-index. For author metrics, venue scores correlate with influential citations~\cite{Valenzuela2015IdentifyingMC} or h-index about as well as such measures correlate against each other or raw citation counts (see Table~\ref{tab:authorlevel}).  For venue metrics, venue scores correlate with h-index (0.64) and citations (0.61) nearly as well as citations correlate with h-index (0.66). For university metrics, venue scores correlate as well with measures of peer assessment as citation-based metrics do~\cite{ScholarRank}. 

As h-index and citation counts have their flaws, obtaining perfect correlation is not necessarily a desirable goal. Instead, these strong correlations serve as evidence for the viability of venue scores. 

Venue scores have been shown to be robust against hyperparameter choices (Tables~\ref{tab:normalizations},~\ref{tab:authormodels},~\ref{tab:combining},~\ref{tab:temporalresults}). Even venue scores produced from completely different data sources tend to look very similar (Table~\ref{tab:corrtable}). Additionally, venue scores can naturally capture the variation of conference quality over time (Figures~\ref{fig:teaser},~\ref{fig:adjust}). 

As with any inductive method, venue scores are data-driven and will be subject to past biases. For example, venue scores can clearly be biased by hiring practices, pay inequality and NSF funding priorities. As these are the supervising metrics, bias in those datasets will be encoded in our results. For example, we found that the faculty hiring metric prioritized Theoretical Computer Science, while using NSF awards prioritized Robotics. The faculty classification task may devalue publishing areas where candidates pursue industry jobs, while the NSF grant regression task may devalue areas with smaller capital requirements. By using large datasets and combining multiple metrics in a single model (Section~\ref{sec:combining}), the final model could reduce the biases in any individual dataset. 

Each of our metrics of interest has an inherent bias in timescale, which our temporal normalization tries to correct for, but likely does an incomplete job of. Salaries are often higher for senior faculty. NSF Awards can have a long response time and a preferences towards established researchers. Faculty classification prioritizes the the productive years of existing faculty. Additionally, faculty hiring as a metric will have a bias towards work from prestigious universities~\cite{clauset2015systematic} and their venue preferences. Some of these issues also exist in citation metrics, and may be why our uncorrected models correlated better with them (Table~\ref{tab:normalizations}).  

\begin{figure*}[ht!]
  \centering
  \includegraphics[width=\linewidth]{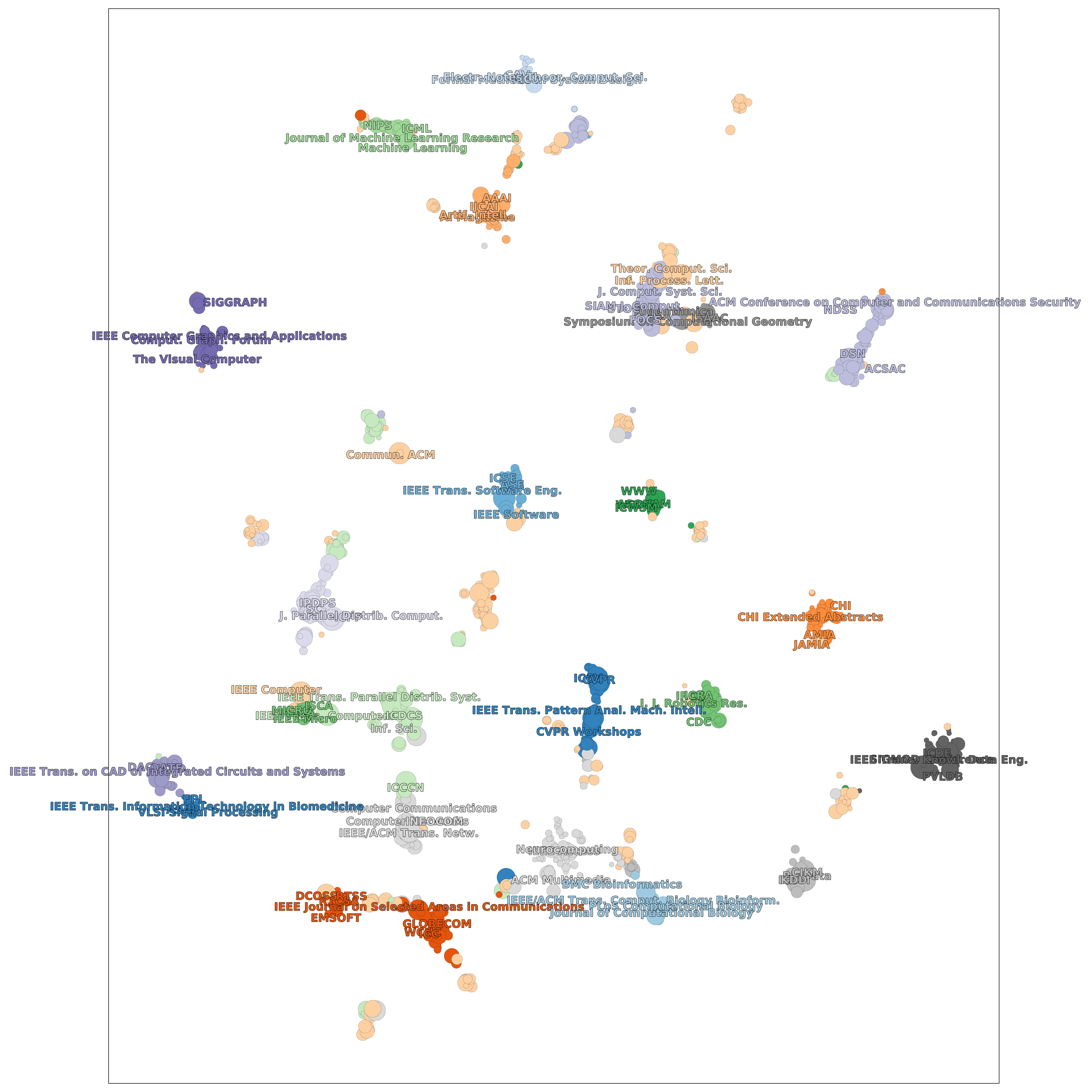} %image19
  \caption{Automatic clustering for venues in Computer Science, the largest venues in each cluster are labeled. }
  \Description{Automatic clustering for venues in Computer Science}
    \label{fig:guide}
\end{figure*}

\begin{figure}
  \centering
  \includegraphics[width=\linewidth]{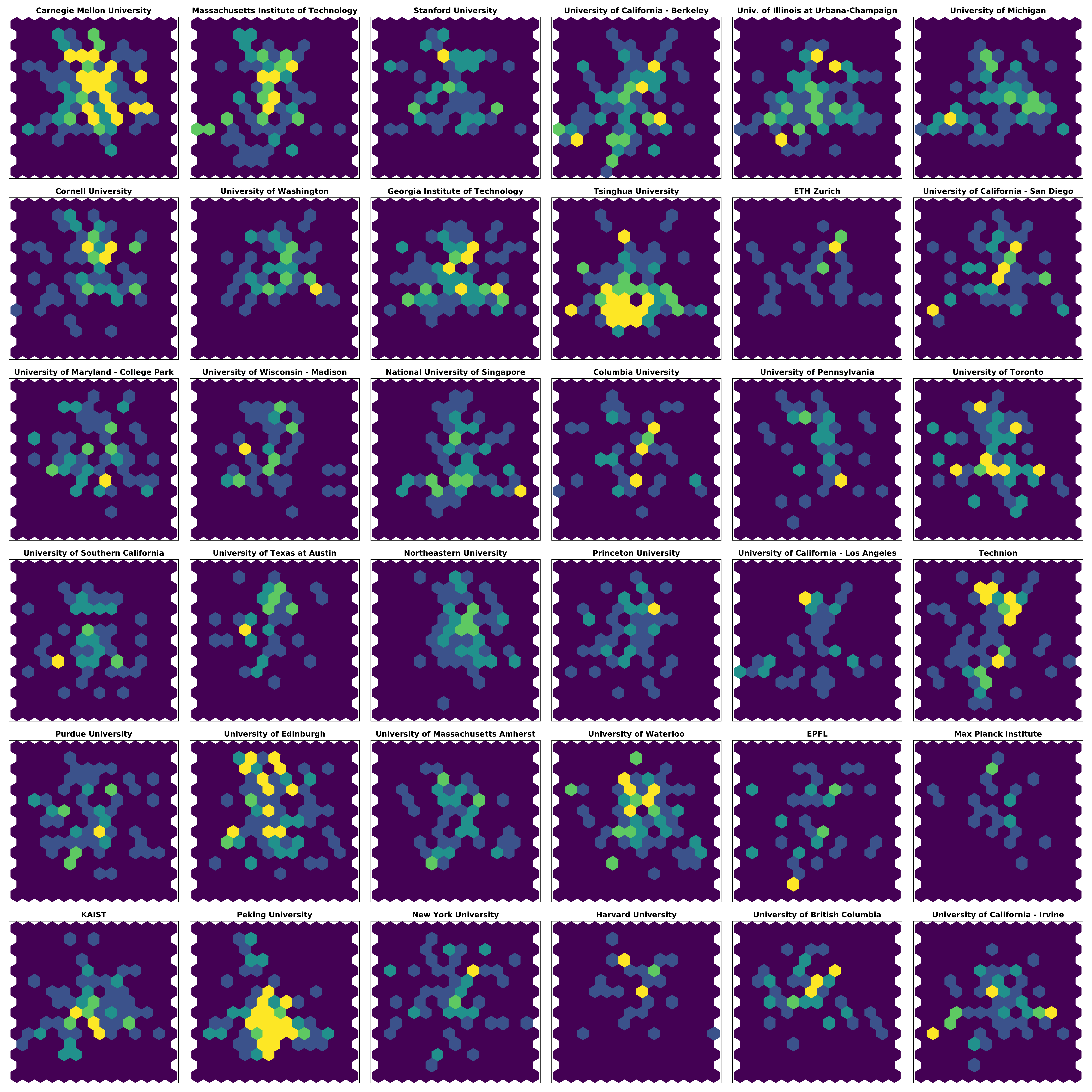}
  \caption{Heatmap showing the differences in research focus across different universities. Figure~\ref{fig:guide} can be used as a guide.}
  \Description{Different university's fingerprints}
  \label{fig:heatmap}
\end{figure}
\begin{figure}
  \centering  
  \includegraphics[width=\linewidth]{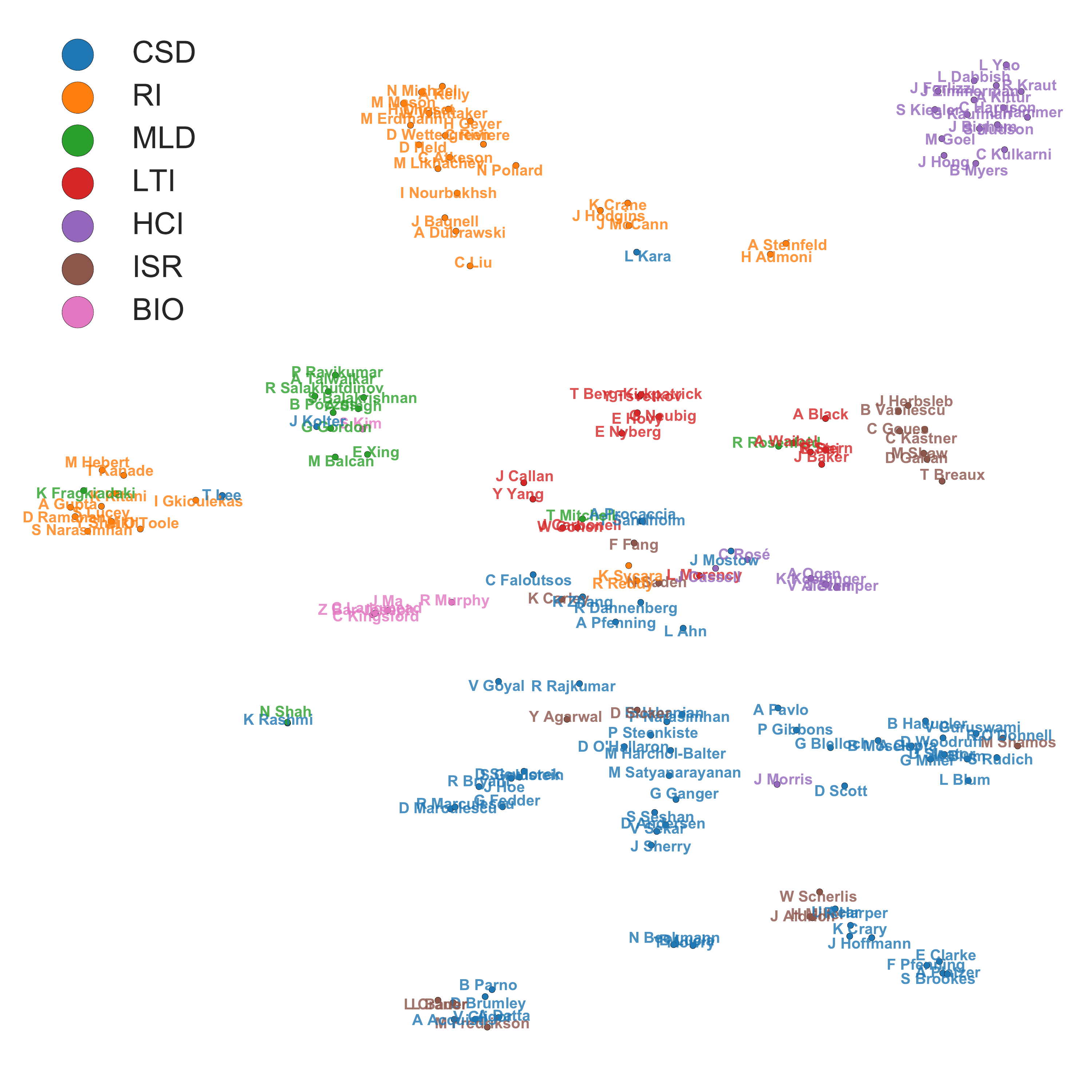}
  \caption{An embedding of Carnegie Mellon University's School of Computer Science with colors indicating sub-departments. For example, the Robotics Institute (RI) has clear clusters for Robotics, Graphics, CV and HRI. }
  \label{fig:cmuscs}
\end{figure}
\section{Similarity Metrics}\label{sec:lda}
While the previous sections of this paper have focused on evaluation, the same dataset can be used to organize venues into groups. For organization, we use a much smaller dataset, using data since 2005 and only evaluating the 1,155 venues that have at least 20 R1 universities with faculty publishing in them. We then build the venue $\times$ author matrix, counting the number of papers each that author published in each venue. Performing a $d$-dimensional Latent Dirichlet Allocation~\cite{blei2003latent}, we obtain a 50-dimensional vector representing each conference in a meaningful way.

These vectors can then be clustered~\cite{arthur2007k} to produce automatic categories for each conference. These high dimensional vectors can also be  embedded~\cite{van2008visualizing} into two dimensions to produce a visual map of Computer Science. See Figure ~\ref{fig:guide} for our result. These clusters represent natural categories in Computer Science. For example, it is easy to see groups that could be called Theory, Artificial Intelligence, Machine Learning, Graphics, Vision, Parallel Computing, Software Engineering, Human-Computer Interaction, among many others.

While some clusters are distinct and repeatable, others are not. When datasets contain challenging cases, ideal clustering can be hard to estimate~\cite{ClusteringIsHard}. Using silhouette scores~\cite{ROUSSEEUW198753}, we can estimate how many natural clusters of publishing exist in Computer Science. In our experiments, silhouette scores were maximized with 20 to 28 clusters. As the clustering process is stochastic, we were unable to determine the optimal cluster number with statistical significance.

%In this case, there exist many nontrivial questions of where to draw the line between fields: does medical imaging belong to computer vision? do the EE and CS communities of distributed systems deserve to be separated?, etc. 

%When looking for six clusters in the DBLP dataset, we found four robust clusters: Design Automation, Data Mining, Networking and Distributed Computing. The remaining areas (PL, OS, NLP, Graphics, Software Engineering, AI) we may call a \textit{core CS} cluster. In one run, Computer Vision became a sixth, separating from core CS. In another run, CS Theory separated and became the sixth cluster. 

By embedding each author with the weighted average of their publication's vectors, we can also obtain a \textit{fingerprint} that shows which areas of Computer Science each university focuses on. See Figure ~\ref{fig:heatmap} for an example of such fingerprints for many top departments. The same clustering method can be used to analyze the focus areas of a single department, for an example see Figure~\ref{fig:cmuscs}.

%\section{Applications}
%The simplicity of venue scores allows for the development of simple queries for other applications. While citation-based metrics exhibit a lag time, venue scores for authors can be updated whenever a new paper is published. Additionally, we can compare the career arcs of several accomplished academics, even if their their disciplines and era are vastly different, as in Figure \ref{fig:fame}.
%For example, the responsiveness of this ranking system, combined with the organization strategy described in Section~\ref{sec:lda}, make this framework useful for evaluating recent PhD graduates. For example, one can screen the 1.7M authors in DBLP for candidates with a short publishing history (in terms of years), and sort them by value to identify promising young researchers. These researchers, using their  similarity metrics, can be paired with existing faculty members who publish in similar areas. 

\section{Conclusion}
We have presented a method for ranking and organizing a scholarly field based only on simple publication information- namely a list of papers, each labeled with only their published venue, authors, and year. By regressing venue scores from metrics of interest, one obtains a plausible set of venue scores. These scores can be compared across sub-fields and aggregated into author-level and institution-level metrics. The scores provided by this system, and their resulting rankings, correlate highly with other established metrics. As this system is based on easily obtainable, publicly available data, it is transparent and reproducible. Our method builds on simple techniques and demonstrates that their application to large-scale data can produce surprisingly robust and useful tools for scientometric analysis.

%\begin{itemize}
%\item {\verb|anonymous,review|}: Suitable for a ``double-blind'' conference submission. Anonymizes the work and %includes line numbers. Use with the \verb|\acmSubmissionID| command to print the submission's unique ID on each %page of the work.
%\item{\verb|authorversion|}: Produces a version of the work suitable for posting by the author.
%\item{\verb|screen|}: Produces colored hyperlinks.
%\end{itemize}
\begin{acks}
Martial Hebert suggested the use of correlation metrics as a technique for quantitative evaluation, thereby providing the framework for every table of results in this paper. Emery Berger developed CSRankings~\cite{CSRankings}, which was highly influential in the design and implementation of this project. Joseph Sill~\cite{sill2010improved}, Wayne Winston, Jeff Sagarin and Dan Rosenbaum developed Adjusted Plus-Minus, a sports analytics technique that partially inspired this work. Kevinjeet Gill was unrelenting in advocating for a year-by-year regression model to avoid sampling and quantization artifacts. 
\end{acks}

% The next two lines define the bibliography style to be used, and the bibliography file.
\clearpage
\bibliographystyle{ACM-Reference-Format}
\bibliography{sample-base}

% If your work has an appendix, this is the place to put it.
\appendix

\section{Credit Assignment}
In order to address issues of collinearity raised by having authors who publish papers together, we wanted to solve a credit assignment problem. We adapted a well-known method for addressing this by adapting regularized plus minus~\cite{sill2010improved} from the sports analytics literature. In our case, we simply regress each publications values from its authors, as shown below. 

\begin{displaymath} \label{eq:apm}
\bordermatrix{& \text{author}_1 & \text{author}_2 &\ldots & \text{author}_n\cr
                \text{paper}_1 & 1 &  1  & \ldots & 0\cr
                \text{paper}_2& 1  &  0 & \ldots & 0\cr
                \vdots& \vdots & \vdots & \ddots & \vdots\cr
                \text{paper}_m& 0  &   1       &\ldots & 1} 
\begin{pmatrix}
x_1 \\ x_2 \\ \vdots \\ x_n 
\end{pmatrix} = \begin{pmatrix}
\text{score}_1 \\ \text{score}_2 \\ \vdots \\ \text{score}_n
\end{pmatrix} 
\end{displaymath}

This technique produced scores that correlated highly with total value scores. While this may be valuable for understanding an individual's contribution, but we were unable to find an evaluation for this method. The form of this model that considers n-wise or pairwise relationships was also considered, but these matrices were too large for us to run trials.

\begin{figure}
  \centering  
  \includegraphics[width=\linewidth]{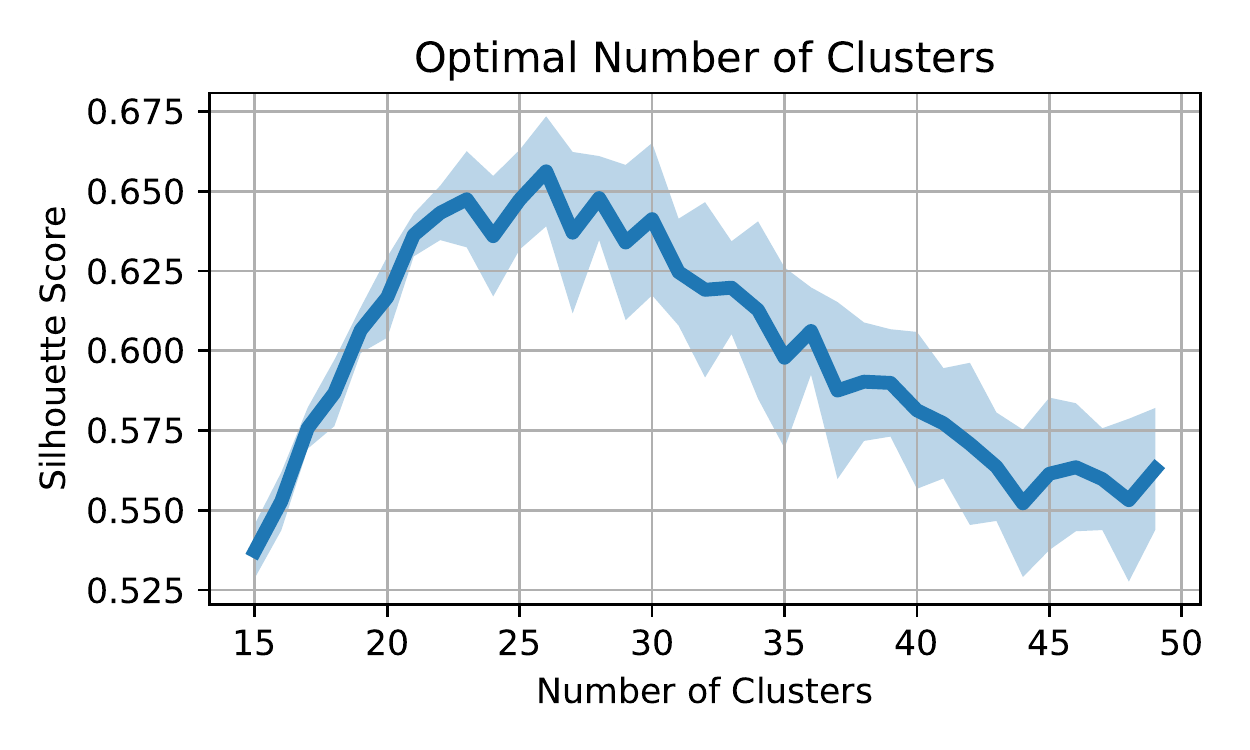}
  \caption{Curve showing the natural number of clusters in the dataset using R1 authors}
  \label{fig:cluster_curve}
\end{figure}

\section{Aging Curve}
\begin{figure}
  \centering  
  \includegraphics[width=\linewidth]{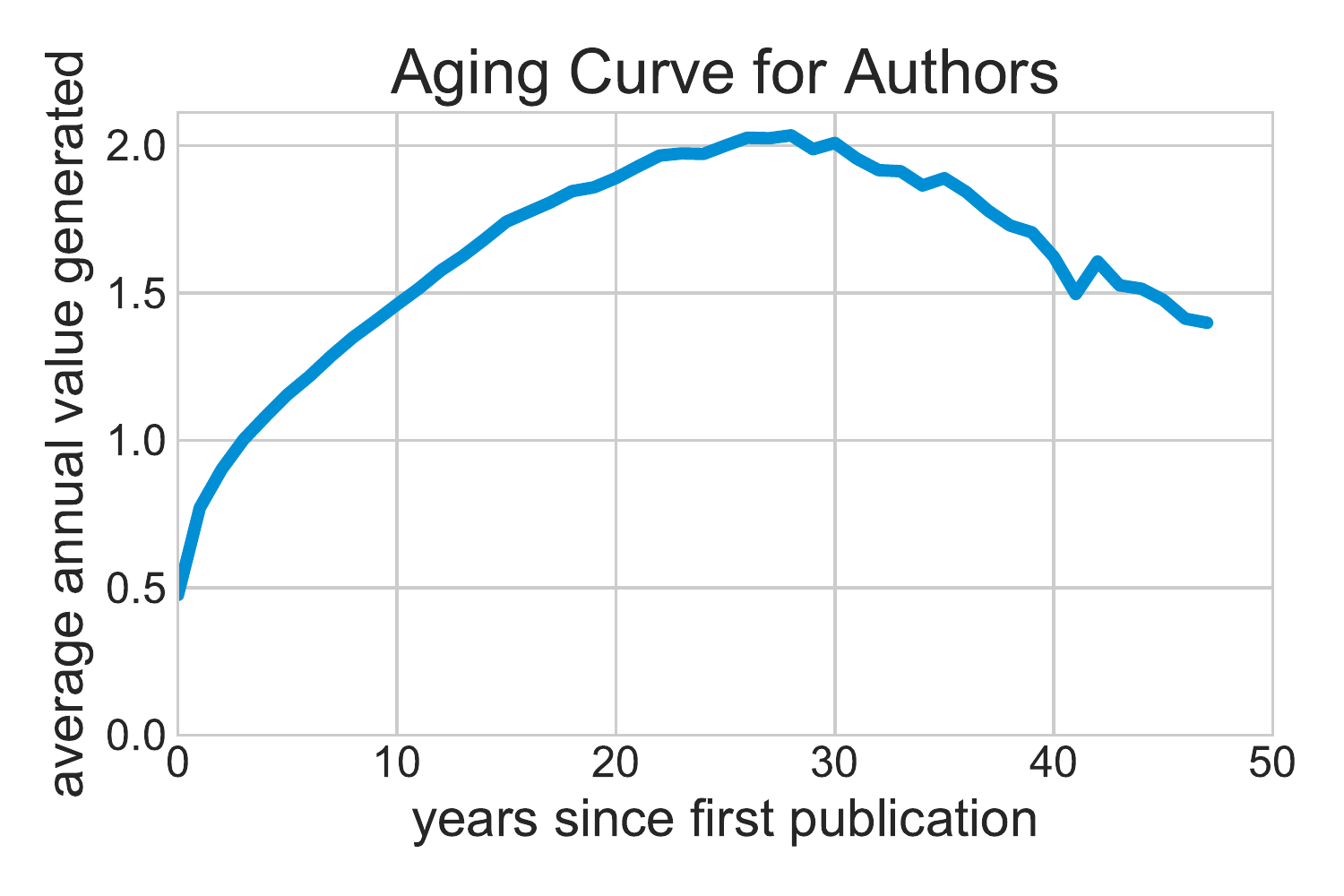}
  \caption{The average productivity of all DBLP authors for that year of their publishing career.}
  \label{fig:age_curve}
\end{figure}

To evaluate if our model makes a sensible prediction over the timescale of a scholar's career, we built a model to see what an average academic career looks like, given that the author is still publishing in those years. See Figure~\ref{fig:age_curve}. Our model suggests a rise in productivity for the first 20 years of one's publishing history, and then a steady decline. 

\begin{figure*}
  \centering  
  \includegraphics[width=\linewidth]{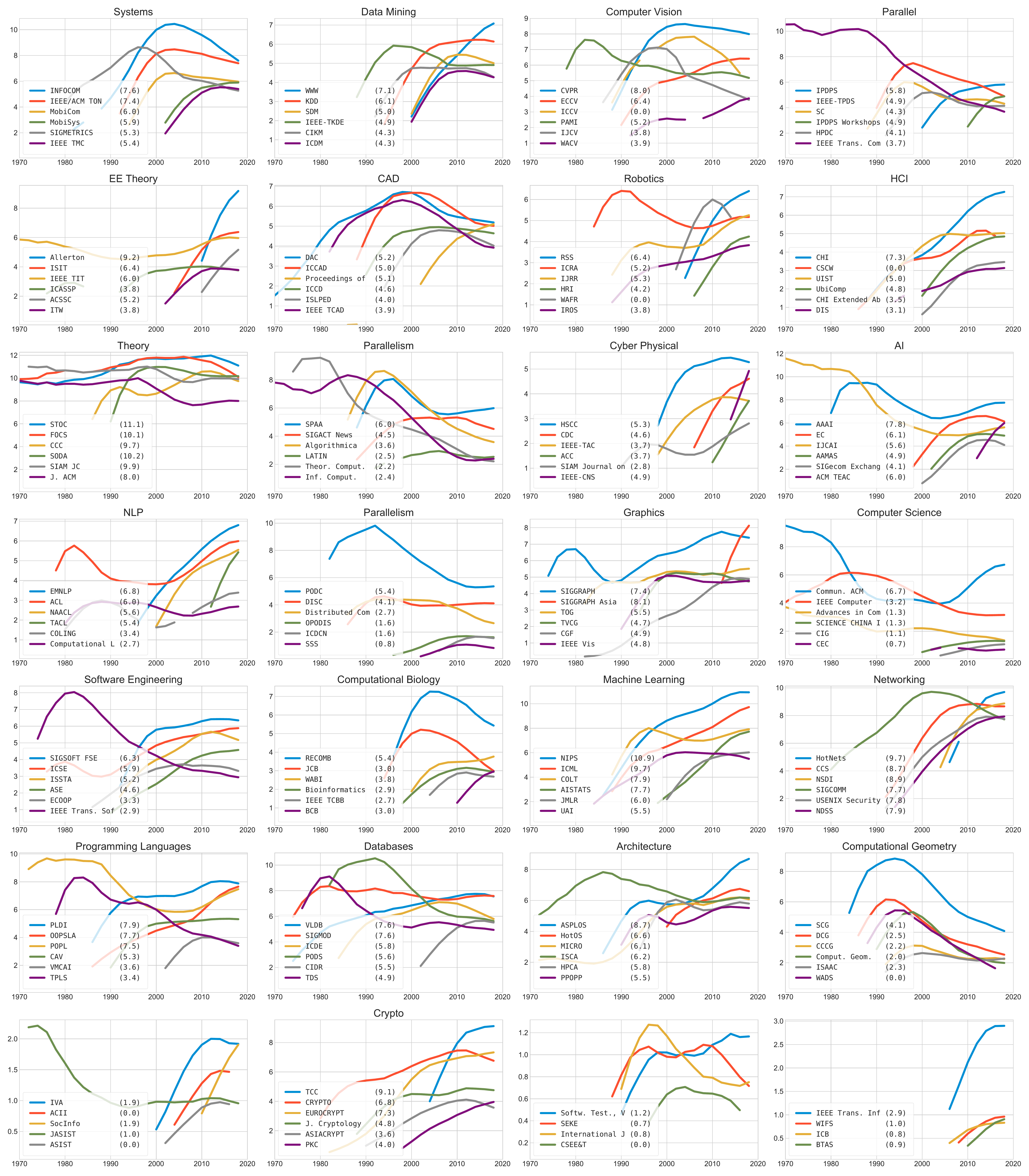}
  \caption{An automatic clustering and rating of the entirity of CS.}
  \label{fig:bigplot}
\end{figure*}
% Please add the following required packages to your document preamble:
% \usepackage{booktabs}
\begin{table*}[]
\centering
\caption{Top 55 Venues}
\label{tab:top-venues}
\begin{tabular}{@{}lll@{}}
\toprule
Name                                   & Score & Size \\ \midrule
STOC                                   & 11.71 & 128  \\
FOCS                                   & 11.04 & 120  \\
NIPS                                   & 10.94 & 484  \\
Conference on Computational Complexity & 10.41 & 53   \\
SODA                                   & 10.17 & 224  \\
SIAM J. Comput.                        & 9.98  & 138  \\
HotNets                                & 9.53  & 46   \\
ICML                                   & 9.46  & 303  \\
TCC                                    & 9.06  & 76   \\
NSDI                                   & 8.76  & 69   \\
CCS                                    & 8.76  & 114  \\
INFOCOM                                & 8.59  & 434  \\
ITCS                                   & 8.56  & 103  \\
Allerton                               & 8.53  & 331  \\
ASPLOS                                 & 8.43  & 48   \\
SIGCOMM                                & 8.31  & 55   \\
CVPR                                   & 8.22  & 606  \\
PLDI                                   & 8.05  & 68   \\
J. ACM                                 & 8.03  & 82   \\
Theory of Computing                    & 7.94  & 29   \\
USENIX Security Symposium              & 7.91  & 62   \\
NDSS                                   & 7.86  & 63   \\
APPROX-RANDOM                          & 7.86  & 74   \\
COLT                                   & 7.78  & 82   \\
VLDB                                   & 7.75  & 185  \\
IEEE/ACM Trans. Netw.                  & 7.74  & 226  \\
AAAI                                   & 7.73  & 403  \\
SIGMOD Conference                      & 7.62  & 98   \\
SIGGRAPH                               & 7.59  & 125  \\
AISTATS                                & 7.52  & 138  \\
OOPSLA                                 & 7.41  & 75   \\
SIGGRAPH Asia                          & 7.38  & 140  \\
CRYPTO                                 & 7.24  & 80   \\
EUROCRYPT                              & 7.24  & 78   \\
POPL                                   & 7.22  & 68   \\
CHI                                    & 7.15  & 431  \\
IEEE SSP                               & 7.11  & 54   \\
WWW                                    & 6.82  & 232  \\
HotOS                                  & 6.75  & 27   \\
EMNLP                                  & 6.61  & 263  \\
EC                                     & 6.60  & 63   \\
Commun. ACM                            & 6.59  & 222  \\
GetMobile                              & 6.50  & 41   \\
IMC                                    & 6.49  & 70   \\
OSDI                                   & 6.46  & 29   \\
ICDE                                   & 6.44  & 183  \\
SIGSOFT FSE                            & 6.42  & 93   \\
ECCV                                   & 6.42  & 252  \\
ICALP                                  & 6.40  & 148  \\
ISIT                                   & 6.28  & 1032 \\
KDD                                    & 6.23  & 240  \\
ICCV                                   & 6.22  & 292  \\
Robotics: Science and Systems          & 6.20  & 88   \\
MICRO                                  & 6.19  & 62   \\
ISCA                                   & 6.18  & 73   \\ \bottomrule
\end{tabular}
\end{table*}
% Please add the following required packages to your document preamble:
% \usepackage{booktabs}
\begin{table*}[]
\centering
\caption{Top 55 Universities}
\label{tab:unis}
\begin{tabular}{@{}lllll@{}}
\toprule
school                                   & authors & papers & venue score & size normed \\ \midrule
Carnegie Mellon University               & 174     & 17275  & 73126       & 14159       \\
University of California - Berkeley      & 108     & 11011  & 51062       & 10884       \\
Massachusetts Institute of Technology    & 108     & 9613   & 47037       & 10026       \\
Univ. of Illinois at Urbana-Champaign    & 103     & 11248  & 44714       & 9627        \\
Technion                                 & 96      & 8795   & 39601       & 8656        \\
Stanford University                      & 69      & 7028   & 36169       & 8513        \\
Georgia Institute of Technology          & 108     & 9337   & 38083       & 8118        \\
Tsinghua University                      & 150     & 14643  & 40662       & 8104        \\
University of California - Los Angeles   & 46      & 6589   & 30368       & 7888        \\
University of Michigan                   & 83      & 7897   & 33666       & 7598        \\
Tel Aviv University                      & 48      & 5468   & 28816       & 7404        \\
University of California - San Diego     & 74      & 6646   & 30836       & 7142        \\
University of Maryland - College Park    & 76      & 7751   & 30795       & 7089        \\
ETH Zurich                               & 39      & 6673   & 26120       & 7081        \\
Cornell University                       & 81      & 5918   & 30278       & 6871        \\
University of Washington                 & 69      & 5727   & 29087       & 6846        \\
University of Southern California        & 58      & 7206   & 26044       & 6387        \\
Columbia University                      & 53      & 5064   & 25251       & 6330        \\
EPFL                                     & 55      & 6927   & 25320       & 6290        \\
Princeton University                     & 47      & 4991   & 24203       & 6252        \\
HKUST                                    & 57      & 6640   & 25134       & 6190        \\
National University of Singapore         & 75      & 7210   & 25123       & 5801        \\
University of Pennsylvania               & 53      & 5340   & 22696       & 5690        \\
University of California - Irvine        & 71      & 6624   & 24070       & 5628        \\
Pennsylvania State University            & 49      & 5668   & 21325       & 5451        \\
Peking University                        & 147     & 11858  & 27050       & 5413        \\
University of Toronto                    & 99      & 5955   & 24759       & 5376        \\
University of Texas at Austin            & 52      & 4485   & 21225       & 5346        \\
University of California - Santa Barbara & 38      & 4698   & 19137       & 5224        \\
University of Waterloo                   & 105     & 7316   & 23781       & 5099        \\
Max Planck Institute                     & 32      & 4307   & 17808       & 5093        \\
Purdue University                        & 71      & 5659   & 21734       & 5082        \\
New York University                      & 67      & 4915   & 20752       & 4918        \\
Rutgers University                       & 63      & 4935   & 20311       & 4884        \\
Nanyang Technological University         & 71      & 9143   & 20827       & 4870        \\
University of Wisconsin - Madison        & 54      & 3988   & 18986       & 4738        \\
University of Massachusetts Amherst      & 57      & 4129   & 19155       & 4717        \\
Hebrew University of Jerusalem           & 42      & 3316   & 17478       & 4647        \\
University of California - Riverside     & 43      & 4230   & 17114       & 4522        \\
Shanghai Jiao Tong University            & 77      & 8220   & 19652       & 4511        \\
Zhejiang University                      & 89      & 7531   & 20232       & 4496        \\
Ohio State University                    & 42      & 4121   & 16740       & 4451        \\
Duke University                          & 23      & 2879   & 13936       & 4385        \\
Chinese Academy of Sciences              & 46      & 5795   & 16497       & 4285        \\
University of Minnesota                  & 43      & 4346   & 15855       & 4190        \\
Northwestern University                  & 49      & 3993   & 16183       & 4137        \\
Stony Brook University                   & 47      & 4565   & 15819       & 4086        \\
University of Edinburgh                  & 115     & 6461   & 19292       & 4058        \\
University of Tokyo                      & 80      & 7524   & 17762       & 4042        \\
University of Oxford                     & 69      & 5549   & 17160       & 4039        \\
Chinese University of Hong Kong          & 32      & 3927   & 13843       & 3959        \\
TU Munich                                & 53      & 6432   & 15521       & 3891        \\
KAIST                                    & 81      & 5591   & 17114       & 3884        \\
Harvard University                       & 34      & 2471   & 13640       & 3837        \\
Imperial College London                  & 65      & 6622   & 15834       & 3779        \\ \bottomrule
\end{tabular}
\end{table*}

% Please add the following required packages to your document preamble:
% \usepackage{booktabs}
\begin{table*}[]
\centering
\caption{Top 50 Authors}
\label{tab:authors}
\begin{tabular}{@{}ll@{}}
\toprule
Name                               & Score \\ \midrule
Philip S. Yu                       & 4354  \\
H. Vincent Poor                    & 3873  \\
Kang G. Shin                       & 3540  \\
Jiawei Han 0001                    & 3143  \\
Micha Sharir                       & 3086  \\
Thomas S. Huang                    & 2992  \\
Don Towsley                        & 2913  \\
Xuemin Shen                        & 2755  \\
Luc J. Van Gool                    & 2700  \\
Noga Alon                          & 2596  \\
Christos H. Papadimitriou          & 2531  \\
Leonidas J. Guibas                 & 2525  \\
Mahmut T. Kandemir                 & 2467  \\
Jie Wu 0001                        & 2452  \\
Wen Gao 0001                       & 2416  \\
Shuicheng Yan                      & 2373  \\
Rama Chellappa                     & 2286  \\
Alberto L. Sangiovanni-Vincentelli & 2266  \\
Georgios B. Giannakis              & 2266  \\
Yunhao Liu                         & 2257  \\
Mohamed-Slim Alouini               & 2238  \\
Michael I. Jordan                  & 2213  \\
Christos Faloutsos                 & 2173  \\
Massoud Pedram                     & 2153  \\
Dacheng Tao                        & 2149  \\
Hans-Peter Seidel                  & 2145  \\
Luca Benini                        & 2139  \\
Moshe Y. Vardi                     & 2092  \\
Lajos Hanzo                        & 2081  \\
Yishay Mansour                     & 2073  \\
Sudhakar M. Reddy                  & 2070  \\
Sajal K. Das                       & 2045  \\
Pankaj K. Agarwal                  & 2042  \\
Avi Wigderson                      & 2024  \\
Robert E. Tarjan                   & 2019  \\
Irith Pomeranz                     & 2003  \\
Elisa Bertino                      & 2003  \\
Zhu Han                            & 1992  \\
Shlomo Shamai                      & 1976  \\
David Peleg                        & 1974  \\
Victor C. M. Leung                 & 1966  \\
Kaushik Roy 0001                   & 1949  \\
Xiaoou Tang                        & 1938  \\
Hector Garcia-Molina               & 1923  \\
Larry S. Davis                     & 1918  \\
Rafail Ostrovsky                   & 1895  \\
Eric P. Xing                       & 1881  \\
Ness B. Shroff                     & 1879  \\
Krishnendu Chakrabarty             & 1868  \\
Hai Jin 0001                       & 1867  \\ \bottomrule
\end{tabular}
\end{table*}

\end{document}